\documentclass{egpubl}
\usepackage{egsgp2026}
 
\SpecialIssueSubmission    %

\CGFccby

\usepackage[T1]{fontenc}
\usepackage{dfadobe}
\ifpdf \PassOptionsToPackage{pdftex}{graphicx}
\else \PassOptionsToPackage{dvips}{graphicx} \fi
\usepackage[export]{adjustbox}
\usepackage{booktabs}

\usepackage{cite}  %
\BibtexOrBiblatex
\electronicVersion
\PrintedOrElectronic
\usepackage{graphicx}
\usepackage{import}
\ifpdf \pdfcompresslevel=9 \fi

\usepackage{egweblnk}

\title{Surface Quadrilateral Meshing from Integrable Odeco Fields}

\author[M. Couplet \& A. Chemin \& D. Bommes \& Edward Chien]
{\parbox{\textwidth}{\centering
   M. Couplet\thanks{Corresponding author}$^{1}$\orcid{0009-0008-9175-425X},
   A. Chemin$^{2}$\orcid{0000-0002-5500-7715}, 
   D. Bommes$^{3}$\orcid{0000-0002-3190-1341}, 
   E. Chien$^{1}$\orcid{0000-0001-5084-7638}
         }
         \\
 {\parbox{\textwidth}{\centering
   $^1$Boston University, $^2$Université catholique de Louvain,
   $^3$University of Bern
        }
 }
}

\newcommand{\R}[0]{\mathbb{R}}

\usepackage{mystyle}
\usepackage{mathtools}

\begin{document}

 \teaser{
  \centering
  \def\svgwidth{\linewidth}
\begingroup%
  \makeatletter%
  \providecommand\color[2][]{%
    \errmessage{(Inkscape) Color is used for the text in Inkscape, but the package 'color.sty' is not loaded}%
    \renewcommand\color[2][]{}%
  }%
  \providecommand\transparent[1]{%
    \errmessage{(Inkscape) Transparency is used (non-zero) for the text in Inkscape, but the package 'transparent.sty' is not loaded}%
    \renewcommand\transparent[1]{}%
  }%
  \providecommand\rotatebox[2]{#2}%
  \newcommand*\fsize{\dimexpr\f@size pt\relax}%
  \newcommand*\lineheight[1]{\fontsize{\fsize}{#1\fsize}\selectfont}%
  \ifx\svgwidth\undefined%
    \setlength{\unitlength}{592.21983422bp}%
    \ifx\svgscale\undefined%
      \relax%
    \else%
      \setlength{\unitlength}{\unitlength * \real{\svgscale}}%
    \fi%
  \else%
    \setlength{\unitlength}{\svgwidth}%
  \fi%
  \global\let\svgwidth\undefined%
  \global\let\svgscale\undefined%
  \makeatother%
  \begin{picture}(1,0.25004548)%
    \lineheight{1}%
    \setlength\tabcolsep{0pt}%
    \put(0,0){\includegraphics[width=\unitlength,page=1]{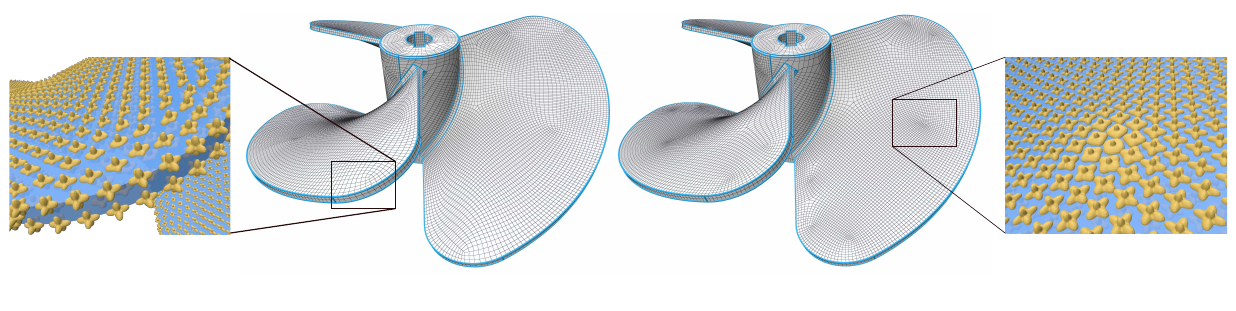}}%
    \put(0.34080806,0.01209302){\color[rgb]{0,0.06666667,0.16862745}\makebox(0,0)[t]{\smash{\begin{tabular}[t]{c}Minimize \textbf{area} distortion\end{tabular}}}}%
    \put(0.63640979,0.01198244){\color[rgb]{0,0.06666667,0.16862745}\makebox(0,0)[t]{\smash{\begin{tabular}[t]{c}Minimize \textbf{angle} distortion\end{tabular}}}}%
    \put(0.90511631,0.04406564){\color[rgb]{0,0.06666667,0.16862745}\makebox(0,0)[t]{\smash{\begin{tabular}[t]{c}Integrable\\odeco field\end{tabular}}}}%
    \put(0.09750539,0.04345756){\color[rgb]{0,0.06666667,0.16862745}\makebox(0,0)[t]{\smash{\begin{tabular}[t]{c}Integrable\\odeco field\end{tabular}}}}%
  \end{picture}%
\endgroup%

  \caption{Our optimization framework utilizes normal-aligned 3D odeco tensors to produce integrable frame fields suitable for parameterization-based generation of anisotropic quad meshes. Our framework also accommodates area- and angle-distortion-miniziming energies. Our method jointly optimizes for singularity positions and integrability, allowing the frames to stray from odeco in the vicinity of naturally arising singularities.}
 \label{fig:teaser}
}

\maketitle
\begin{abstract}
   We present a method for generating orthogonal quadrilateral meshes subject to user-defined feature alignment and sizing constraints. The approach relies on computing integrable orthogonal frame fields, whose symmetries are implicitly represented using orthogonally decomposable (odeco) tensors. We extend the existing 2D odeco integrability formulation to the 3D setting, and define the useful energies in a finite element approach. Our frame fields are shear-free (orthogonal) by construction, and we provide terms to minimize area and/or angle distortion. The optimization naturally creates and places singularities to achieve integrability, obviating the need for user placement or greedy iterative methods. We validate the method on both smooth surfaces and feature-rich CAD models. Compared to previous works on integrable frame fields, we offer better performance in the presence of mesh sizing constraints and achieve lower distortion metrics.

\printccsdesc   
\end{abstract}  
\section{Introduction}

The problem of anisotropic quad meshing of curved surfaces has broad application and relevance to a wide number of use cases, e.g., FEM on rectangular meshes \cite{ArnoldBoffiFalk2005}, discrete nets and principal stress networks for physical construction \cite{Pottmann07Architectural,Pellis2018Aligning}, and textile modelling via Chebyshev nets \cite{SagemanFurnas19Chebyshev}. Meshes aligned to lines of principal curvature are also used to guide the creation of such meshes for the purposes of illustration and visualization \cite{Hertzmann00Illustrating}, and for subdivision modeling \cite{ExosideQuadRemesher}. 

Befitting such a central problem, there is an extensive literature aimed at generating semi-regular quad meshes in the presence of alignment and sizing constraints. The most popular sort of pipeline for producing such meshes involves two main steps: (1) a \textit{seamless parameterization} is generated and (2) integer rounding and mesh extraction is performed to achieve the final mesh. For step (1), the classic approach is to first generate a smooth \textit{cross field} and then modify this (unit-length) field to achieve an integrable result that can be used to produce a seamless parameterization. A key determination in this first step is the global topology of the quad mesh, consisting of mesh singularity placements. These placements are reflected in singularities of the initial cross field, whose placements may be suboptimal for a full seamless parameterization, in light of the fact that the field is non-integrable.

In our method, we present a new optimization formulation for step (1) that jointly optimizes for singularity positions and integrability of an orthogonal \emph{frame field} (with variable-length frame vectors) suitable for generation of a seamless parameterization. Such frame fields are represented with \textit{orthogonally decomposable tensors} as introduced in \cite{Robeva16Orthogonal} and applied in \cite{Palmer20Algebraic}. Our work extends the framework of \cite{Couplet26Size}, which utilized 2D odeco tensors in the planar setting, to curved surfaces in 3D with the use of normal-aligned 3D odeco tensors. The 2D integrability energy present there is generalized to a 3D integrability energy that captures intrinsic curl on the 3D surface, and alignment and sizing of fields can again be imposed with simple linear constraints. As seen there, the combination of an integrability and ``odeco-ness'' energy allows for frame field singularities to arise naturally where violation of the latter allows for better global satisfaction of the former.

The use of normal-aligned odeco tensors also generalizes the use of normal-aligned octahedral frames in \cite{zhang20octahedral} to generate cross fields on curved surfaces. The additional variables of frame vector lengths allow consideration of integrability, and our results also exhibit a natural alignment with sharp creases and principal curvature alignment in areas of high curvature. Lastly, we note that we also formulate effective area- and angle-preserving distortion energies within our tensor optimization framework. 

To validate our approach, we run our method on a selection of feature-rich CAD models from the MAMBO dataset \cite{mambo_dataset}, as well as a handful of smooth models from the dataset of \cite{Myles:2014:RFG} (sans features). On the MAMBO models, we compare to the method of \cite{Diamanti2015Integrable}, which is the only publicly available method we are aware of that also attempts to produce frame fields that are also integrable in a single optimization. Empirically, we find that our method leads to better global placement of singularities leading to more orthogonal quad meshes and better satisfaction of sizing constraints. We also compare to \cite{Corman25Rectangular} on a small set of 3 challenging planar domains and one MAMBO model. Their base method assumes singularity placement as input, but we compare to their proposed automatic greedy insertion of singularities via iterative optimization. We find that our method achieves better distortion metrics and mesh symmetry at comparable singularity counts.

\section{Related Work}

The problems of quad meshing, frame field generation, and surface parameterization have been extensively studied, with an associated literature that is too large to comprehensively review here. Thus, we touch on the most relevant aspects and works below and refer readers to some excellent surveys on the topics \cite{bommes13QuadSurvey,Vaxman2016DirectionalSurvey,Floater05Parameterization,sheffer_mesh_2007} for additional coverage.

Our method aims to generate integrable, orthogonal frame fields that integrate to locally injective seamless parameterizations. Moreover, the method handles both feature alignment and element sizing constraints, and jointly optimizes for both singularity placement and distortion measures suitable for generation of anisotropic quad meshes. Of the many works that consider frame/cross field design and seamless parameterization for quad meshing, most have some, but not all of these aspects. We restrict our discussion to methods that fall within the domain of training-free field-guided meshing, leaving out methods based on learning, e.g., \cite{Hao2024MeshtronHA}, advancing fronts, e.g., \cite{MERHOF2007860}, triangle merging, e.g.,  \cite{Remacle2013Frontal}, and principal curvature tracing, e.g., \cite{alliez2003anisotropic}, amongst others.

\vspace{0.1cm}
\noindent \textit{Singularity Placement.} Most methods assume that singularities are given by those of an input cross field derived from smoothed principal curvature directions, e.g., \cite{Kalberer07QuadCover}, or a field optimization that necessarily prioritizes smoothness over integrability (as the cross field is the same size everywhere), e.g., \cite{Bommes09MIQ}. Within the setting of conformal parameterization in the presence of cones (for generation of isotropic quad meshes), there have been many sophisticated works aimed at optimal cone placement to minimize area distortion and related isotropic distortion measures. Some are based on the use of Gaussian curvature as a signal \cite{BenChen2008Conformal}, or via incremental flattening and concentration of this signal \cite{Myles2012Incremental,Myles2013Constrained}. Others formulate more complicated optimizations, that are attacked with advanced optimization techniques like Fenchel-Rockafellar duality \cite{Soliman:2018:OCS} and Douglas-Rachford splitting \cite{Li:2022:SparseCones}. Another recent method that incorporates singularity placement for cross fields is that of \cite{palmer2024lifting}, which selects smooth cross fields as minimal area sections of a circle bundle (but clearly does not incorporate an integrability criterion). 

In contrast, our method achieves reasonable singularity placement while considering integrability and allowing for non-conformal distortion. In particular, our work allows for rectangular parameterizations, for which the coordinate axes get mapped to orthogonal directions. These were considered in the recent work \cite{Corman25Rectangular} which does not jointly optimize for singularity positions. They do posit a greedy, iterative procedure reminiscent of \cite{Myles2012Incremental,Myles2013Constrained} if automatic cone placement is desired. Also notable is the method of \cite{Diamanti2015Integrable}, which achieves integrable PolyVector fields, and also allows for joint optimization of singularity positions and frame fields, though orthogonality of the frames is only enforced softly.

\vspace{0.1cm}
\noindent \textit{Integrability.} As noted above, most methods optimize first for a non-integrable cross field and then modify the frames to achieve integrability through one of three approaches. Most common perhaps is the direct or indirect use of a Helmholtz-Hodge decomposition to extract the curl-free part of the initial field \cite{Kalberer07QuadCover,Bommes09MIQ}. This is done implicitly in any method that utilizes a Poisson solve to find the parameterization that most closely matches the initial cross field. The second common approach is a rescaling of the cross field axes whether isotropic \cite{Ray06Periodic} or anisotropic \cite{Zhang10Wave,Simons24Anisotropy} that aims to achieve integrability. This method fixes the directionality of the input cross field and can lead to high distortion and inherit the poor singularity positioning of the initialization. Lastly, there are recent works that express integrability in other frameworks, like the moving frames setting \cite{Coiffier23Frames,Corman25Rectangular} and PolyVector fields \cite{Diamanti2015Integrable,SagemanFurnas19Chebyshev}. Also notable is the work of \cite{Jezdimirovic24Integrable}, which achieves integrable frame fields on surfaces, but is also constrained to the singularities of an input cross field.

\vspace{0.1cm}
\noindent \textit{Orthogonality and Distortion Control.} There is a large body of work on sophisticated methods for optimizing distortion measures that favor isotropic (conformal) distortion, e.g., \cite{shtengel2017composite,Rabinovich2017SLIM,chien2016Metrics}, but there are few that use distortion measures that allow for anisotropic rectangular distortions. One exception to this is \cite{Levi2023Shear}, which formulates a bespoke shear energy, but thus only enforces orthogonality in a soft fashion. Similarly, \cite{Diamanti2015Integrable} motivate orthogonality via a parameter $s \in [0,1]$. The only method which maintains orthogonality in a strict fashion is the work of \cite{Corman25Rectangular}, which satisfies it by construction in their optimization formulation. Our method achieves orthogonality via an odeco energy, which measures tensor deviation from the odeco variety, where frames are guaranteed to be orthogonal. Empirically, we achieve better orthogonality than \cite{Diamanti2015Integrable}, and the relaxing of this hard orthogonality constraint allows for the joint optimization of singularity positions.

\vspace{0.1cm}
\noindent \textit{Feature Alignment and Sizing Control.} Many works allow for hard or soft feature alignment via an aligned input field, but few allow for the capability to explicitly constrain sizing on boundary constraints. Some notable exceptions are the wave-based \cite{Zhang10Wave}, which only enforce the sizing softly, and metric modification approaches \cite{kovacs2010Anisotropic,jiang2015Metric}, which do not incorporate integrability criteria into their frame field generation. 

\vspace{0.1cm}
\noindent \textit{Local Injectivity.} Many works also focus on criteria for and guarantees of local injectivity through conformal \cite{Gillespie2021Conformal,Campen2021Conformal,fu_inversion-free_2021,Capouellez2024Seamless,Capouellez2025Feature}, harmonic \cite{floater2003one,bright2017HGP}, or more general mappings \cite{lipman2012bounded,garanzha2021foldover,du2022isometric}. The vast majority of these do not target local injectivity of rectangular anisotropic mappings. There are also several notable works that focus on topological guarantees of local injectivity via explicit construction of locally-injective seamless maps \cite{Shen:2022:WCF,Campen2019Seamless,Levi2023Seamless}. However, these all assume specified singularity positions, and usually are wildly distorted and non-orthogonal, relying on subsequent optimization to smooth the mappings. 

At a continuous level, our method enforces positive lengths and integrability for our frames, so is guaranteed to be locally-injective. However, as with \cite{Corman25Rectangular}, it may fail to be injective at a discrete level, and so we use a Poisson method to recover a mapping that is mostly injective. As there, one could also consider giving our frames to a robust parameterization method that can maintain local injectivity, e.g., \cite{Myles:2014:RFG}. 

\section{Background} \label{sec:background}

\subsection{Seamless parametrization} \label{sec:seamless}

\begin{figure}[ht]
	\centering
	\def\svgwidth{.8\linewidth}
	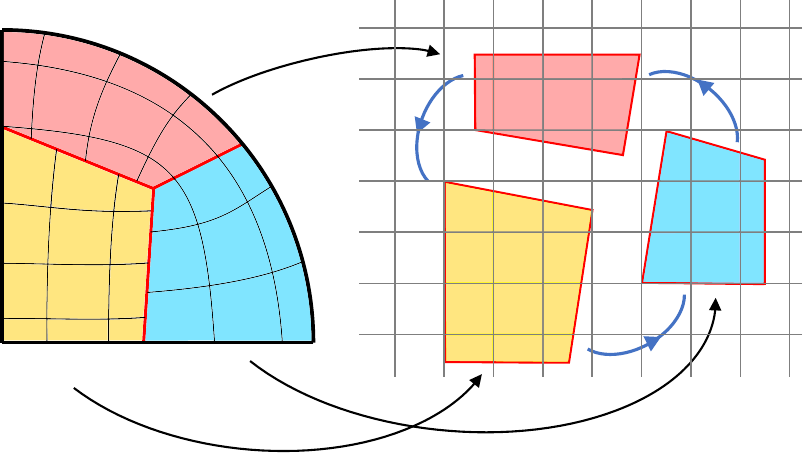
	\caption{Domain $\cal{M}$ is mapped onto a parametric plane by a seamless parametrization $\phi$,
	allowing to map a grid back onto $\cal{M}$.
	}
	\label{fig:seamless-param}
\end{figure}

Let $\cal{M} \subset \bb{R}^3$ be a two-dimensional manifold representing the surface to be meshed.
A \emph{parametrization} $\vb{\phi}:\cal{M}\to\bb{R}^2$ maps the surface to a Cartesian parametric plane;
we sometimes write it as a pair of maps $(\phi^u, \phi^v)$.
By pulling back the integer grid $ (\bb{R}\times\bb{Z}) \cup (\bb{Z}\times\bb{R}) $ from the parametric plane
to the surface, one can locally tesselate the surface into quadrilateral elements.
Because of the topology of the surface or the presence of desired cone points, it is usually impossible to generate a globally continuous parametrization,
and instead one defines a collection of \emph{charts} $\{\phi_i\}$
homeomorphically mapping a partition $M_i \subset \cal{M}$ to subsets $\Omega_i \subset \bb{R}^2$ of the parametric plane.
These maps need to respect a set of conditions to be appropriate for quadrilateral meshing purposes.
First, each chart must be \emph{locally injective} to avoid any inverted elements when pulling back the integer grid onto the surface.
For this property to hold, the Jacobian matrix $\vb{J}_{\phi_i}$ must have positive determinant:
\begin{equation} \label{eq:local_injectivity}
    \det\,\vb{J}_\phi(p) = \det\,\begin{pmatrix} \grad\phi^u \\ \grad\phi^v \end{pmatrix} > 0 \quad \text{for $p\in\cal{M}$.}    
\end{equation}
\revision{$\phi^u$ and $\phi^v$ are the surface gradients of the respective scalar fields $u$ and $v$.}
Charts are glued together by \emph{transition functions}:
if $\phi_i$ and $\phi_j$ are two charts,
then on the intersection of their domains $ M_i \cap M_j$ \revision{(a one-dimensional curve)} there exists
another homeomorphism $g_{i \to j}: \phi_i(M_i \cap M_j) \to \phi_j(M_i \cap M_j)$ with $g_{i \to j} = \phi_j \circ \phi_i^{-1}$.
In order for the mesh to seamlessly stitch between charts, the transition functions need to preserve the Cartesian grid,
which yields the \emph{comformity} condition
\begin{equation} \label{eq:conformity}
    g_{i \to j}(z) = \vb{R}_{i \to j} z + \bm{\tau}_{i\to j} \quad \text{for $z \in \Omega_i \cap \Omega_j$},
\end{equation}
where $\vb{R}_{i\to j}$ is any rotation multiple of $\pi/2$ --
this multiple is denoted $r_{i\to j}$ and is often called a \emph{matching} --
and $\bm{\tau}_{i\to j} \in \bb{R}^2$ is a parametric translation,
\revision{i.e., the offset in parameter space when transitioning from chart $i$ to $j$}.

In the neighborhood of a point $p$, if there is a counterclockwise cycle of charts $\phi_1, \phi_2,\ldots, \phi_N = \phi_1$ about $p$ (as in \cref{fig:seamless-param}) for which the sum of rotation multiples $r_p^\circlearrowleft \coloneq r_{1 \to 2} + r_{2 \to 3} + \ldots +r_{(N-1) \to 1}$ is nonzero, then $p$ is a \textit{singularity} of index $r_p^\circlearrowleft / 4$. Upon quadrangulation such a singular point will become a valence $(4-r_p^\circlearrowleft)$ mesh vertex. The point in the center of Fig. \ref{fig:seamless-param} is of index $+1/4$ and would result in a valence 3 vertex.

A parametrization $\{\phi_i\}$ is said to be \emph{seamless} if it is both locally injective and conforming,
verifying \Cref{eq:local_injectivity,eq:conformity}.
In the presence of feature or boundary curves that the mesh should align to,
we additionally ask that one of the parametric functions $\phi^u$ or $\phi^v$ be constant along the curve.
This is imposed by ensuring that one of the parametrization gradients is orthogonal to the curve;
if $\vb{e}_t$ is a vector tangent to the curve then
\begin{equation} \label{eq:feature_align}
    \grad{\phi^u} \cdot \vb{e}_t = 0 \quad \text{or} \quad \grad{\phi^v} \cdot \vb{e}_t = 0.
\end{equation}

Lastly, for extraction of a quad mesh by pulling back the integer grid, we must require that the parameterization be \textit{integer seamless}. In particular, all singular points $p$ map to integer points, all feature curve points $q$ map into the integer grid, and all parametric translations are integer: 
\begin{gather} \label{eq:int_seamless}
    \phi_i(p) \in \bb{Z} \times \bb{Z} \quad \text{and} \quad \phi_i(q) \in  (\bb{R}\times\bb{Z}) \cup (\bb{Z}\times\bb{R}) \\
    \text{and} \quad \tau_{i \to j} \in \bb{Z}^2.
\end{gather}

\subsection{Integrable frame fields}

The conditions defining a seamless parametrization, \cref{eq:local_injectivity,eq:conformity,eq:feature_align},
can all be expressed as conditions on the parametrization Jacobian $\vb{J}_\phi$.
This is already the case for local injectivity and feature alignment.
For conformity, on the chart boundary we have that $\phi_j = g_{i\to j} \circ \phi_i$,
and taking the Jacobian on both sides gives
\begin{equation}
	\vb{J}_{\phi_j}(p) = \vb{R}_{i \to j} \, \vb{J}_{\phi_i}(p),
	\quad \text{for $p \in M_i \cap M_j$}.
\end{equation}
\begin{figure}[ht]
	\includegraphics[width=\linewidth]{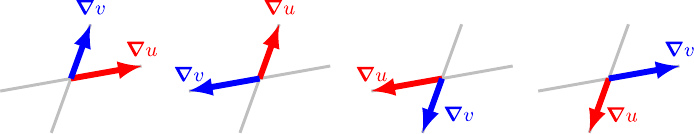}
	\caption{Equivalent gradient pairs $(\grad{u},\grad{v})$
	across the charts of a global seamless parametrization.}
	\label{fig:equiv_frames}
\end{figure}

\begin{figure}[b]
    \centering
    \includegraphics[width=.45\linewidth]{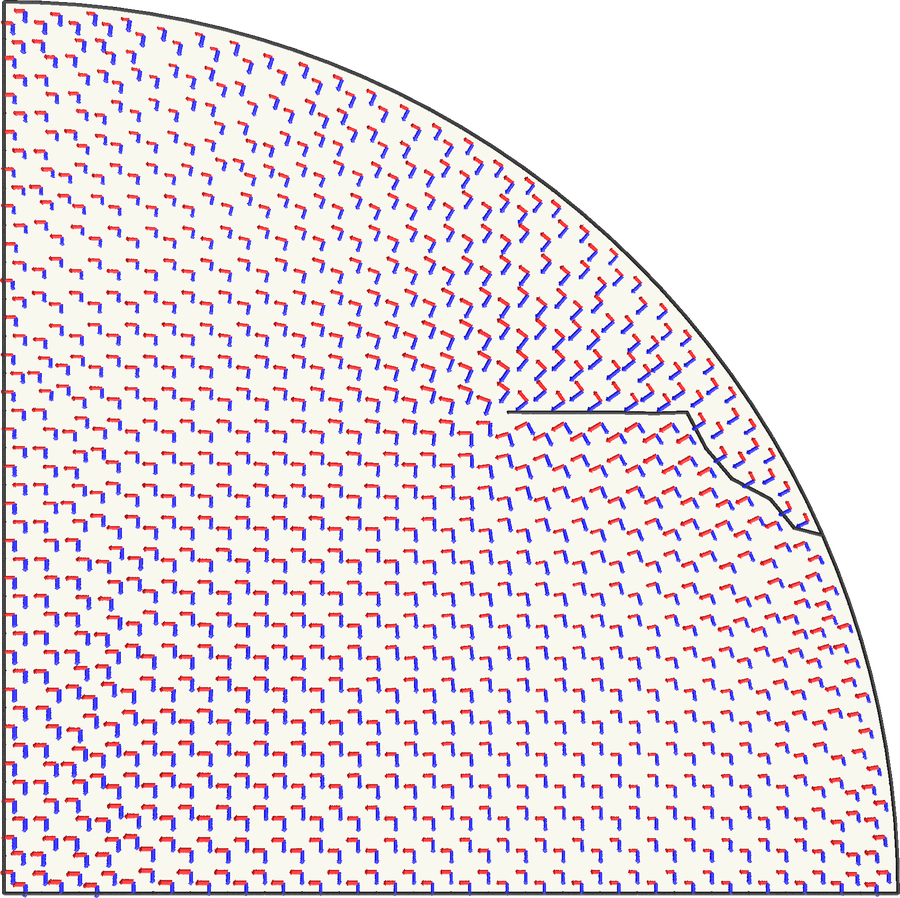}\hspace{1em}
    \includegraphics[width=.45\linewidth]{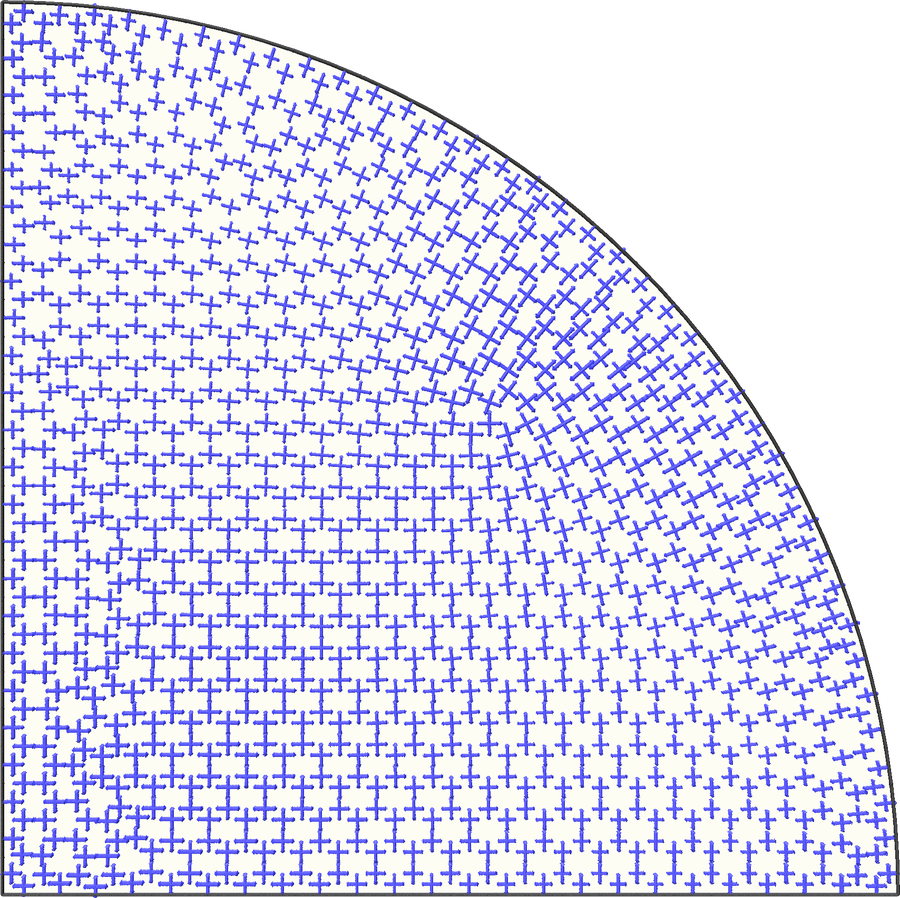}
    \caption{Gradients of a parametrization and corresponding integrable frame field.}
    \label{fig:param-jacobian}
\end{figure}
Because $\vb{R}_{i\to j}$ is a rotation by a multiple of $\pi / 2$, it acts on the rows of $\vb{J}_{\phi_i}$ via permutation (with some signs). Thus, the gradient vectors (rows of $\vb{J}_{\phi_i}$) are not actually rotated, but rather are permuted as shown in Fig. \ref{fig:equiv_frames}. The Jacobian is thus discontinuous across charts where it undergoes a rotation of a multiple of $\pi/2$.
To remove this discontinuity, we can define the equivalence class of $\vb{J}_\phi$ under this group:
\begin{equation} \label{eq:frame_equiv}
    \qty[ \vb{J}_\phi ] \coloneq \qty { \vb{R}^k \, \vb{J}_\phi \ | \ k \in \bb{Z} },
\end{equation}
where $\vb{R}$ performs a $\pi/2$ rotation. 
This equivalence class forms our notion of \emph{frame}.
The corresponding \emph{frame field} $\qty[\vb{J}_\phi](p)$ is then continuous on the whole parametrization,
except at singularities
\revision{(note that the discontinuity is at the singular point,
as sizes of the frame vectors blow up when approaching singularities of valence 3)}. 
This observation justifies computing parametrizations through continuous frame fields,
which do not require integer constraints or cutting the domain into charts.
For a frame field to induce a parametrization, it must be \emph{integrable}.
On any simply-connected domain, a vector field is integrable to a scalar potential if it is \emph{curl-free}.
Let $\vb{F}(p)$ be a frame field as defined by the equivalence class in \cref{eq:frame_equiv}.
At a nonsingular point $p$, a \emph{lifting} of $\vb{F}$ in a neighborhood of $p$ is a pair
of continuous vector fields $\vb{u}(p), \vb{v}(p)$, each being part of frame $\vb{F}(p)$.
Frame field $\vb{F}$ is curl-free at $p$ if any lifting in a neighborhood $\cal{N}(p)$ is curl-free:
\begin{equation} \label{eq:curl-free}
    \curl\vb{u}(p') = \curl\vb{v}(p') = \vb{0}, \quad \text{for $p'\in\cal{N}(p)$}.
\end{equation}
If $\vb{F}$ is curl-free, it is thus the Jacobian of a parametrization $\phi_i$
on any simply-connected chart $M_i$.
Stitching the charts together through the permutations of the frame representation,
$\vb{F}$ is the Jacobian of a global seamless parametrization on the whole domain.

\subsection{Intrinsic vector field curl from extrinsic frames}

Given a vector field $\vb{v}$ tangent to a smooth embedded surface $\cal{S} \subset \R^3$,
it is important to distinguish the intrinsic scalar surface curl $\grad_\cal{S}\times\vb{v}$
from the extrinsic 3D vorticity vector $\curl\vb{\tilde{v}}$ (where $\vb{\tilde{v}}$ is an ambient extension of $\vb{v}$ defined in some neighborhood of $\cal{S}$). If $\vb{n}$ is the surface unit normal,
these quantities are related by
\begin{equation} \label{eq:intrinsic-curl}
    \grad_\cal{S} \times \vb{v} = \vb{n} \cdot (\curl\vb{\tilde{v}}).
\end{equation}
This result requires that $\vb{v} \cdot \vb{n} = 0$ ($\vb{v}$ is a tangent vector field) and is independent of the particular extension $\vb{\tilde{v}}$. This fundamental fact tells us that the normal component of the vorticity is an intrinsic quantity, and is discussed in many classic references, e.g., \cite{AbrahamMarsdenRatiu1988,Frankel2011}.
Thus, we say that $\vb{v}$ is curl-free and integrable on the surface if its 3D vorticity is orthogonal to the surface normal $\vb{n}$.

\begin{figure}[b]
\centering
\includegraphics[width=0.8\linewidth]{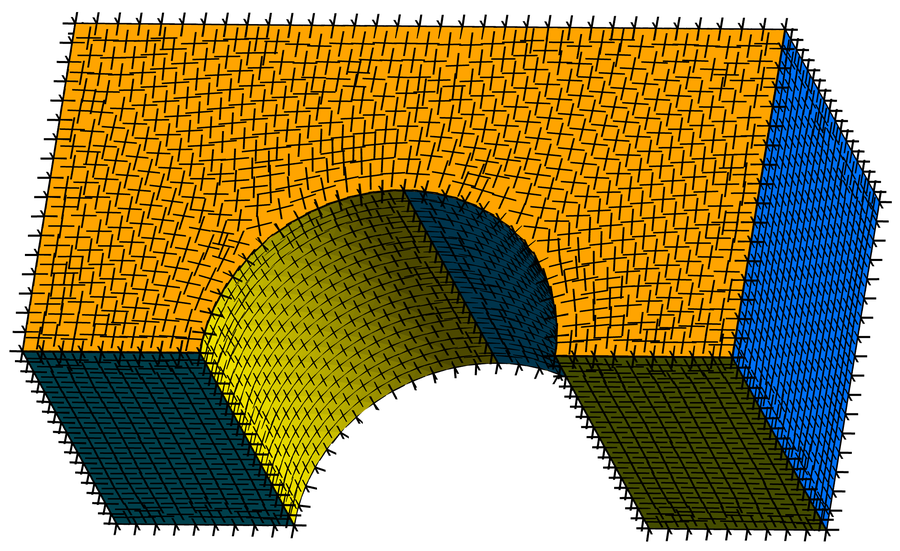}
	\caption{Even with   \emph{no explicit feature alignment constraints}, we still achieve natural frame alignment to sharp creases and principal curvature directions in this result on the B0 model from the MAMBO dataset \cite{mambo_dataset}.}
	\label{fig:natural_alignment}
\end{figure}

We also note here that our use of an extrinsic representation via normal-aligned 3D odeco tensors, as described in \Cref{sec:odeco_representation,sec:alignment}, takes inspiration from the works of \cite{Jakob2015Instant,zhang20octahedral}, both of which use a similar extrinsic frame representation for curved surfaces. As noted in both of these works, the use of such an extrinsic representation allows for natural alignment to sharp creases and principal curvatures in regions of high sectional curvature. Examples of this may be seen in \Cref{fig:natural_alignment} and the last two models of \Cref{fig:area-angle-comparison} (e.g., see arm regions).
\revision{Another advantage of our extrinsic formulation
is that surface curl is computed from 3D curl directly,
as shown later in \cref{sec:odeco_integrability}.
We obviate the need to connect tangent spaces and explicitly account for surface curvature,
making the formulation more straightforward.}

\subsection{Odeco representation} \label{sec:odeco_representation}

Orthogonally decomposable (\emph{odeco}) tensors have been applied in \cite{Palmer20Algebraic}
as a representation of three-dimensional orthogonal frames.
We review their definition and some of their important properties,
before presenting in \Cref{sec:method} how they are exploited in this work.

When looking for a representation of an $n$-dimensional \emph{orthogonal} frame $(\vb{v}_1, \dots, \vb{v}_n)$
that is invariant under permutation and sign changes,
it is tempting to simply consider the $n$-by-$n$ symmetric positive definite matrix
\begin{equation} \label{eq:spdMatrixRep}
    \vb{P} \coloneq \sum_m \lambda_m \vb{u}_m \vb{u}_m^\intercal,
\end{equation}
where $ \lambda_m \coloneq \norm{\vb{v}_m} $ and
$ \vb{u}_m \coloneq \vb{v}_m / \lambda_m $ is the normalized frame vector.
This choice is problematic when some frame vectors have identical lengths,
as they form an eigenspace of $\vb{P}$ of dimension greater than 1
and the information on their individual orientations is lost.
This problem is obviated by turning to a fourth-order tensor representation
\begin{equation} \label{eq:odeco_defn}
    \vb{T} \coloneq \sum_{m=1}^n \lambda_m \vb{u}_m^{\otimes 4}.
\end{equation}
An even order is necessary to ensure invariance under sign changes in the vectors.
This tensor is \emph{fully symmetric}, i.e., $T_{ijkl}$ is invariant under permutation of its 4 indices.
The notion of eigenvalue and eigenvector can be generalized to higher-order tensors as done in \cite{Lim2005SingularVA,Qi2007Eigenvalues}.
A unit vector $\vb{u} \in \bb{R}^n$ is an eigenvector of $\vb{T}$ with eigenvalue $\lambda$
if contracting the tensor three times with $\vb{u}$ results in a scalar multiple:
\begin{equation}
    \vb{T} \cdot \vb{u}^3 = \lambda\vb{u}.
\end{equation}
Hence, the frame vectors and their magnitudes can be interpreted as eigenpairs of their odeco tensor. 

Note that these fourth-order tensor representations do not suffer the same eigenspace degeneracy problems as the second-order symmetric matrix representation \cref{eq:spdMatrixRep}. In particular, if $\vb{u}$ and $\vb{v}$ are both eigenvectors of $\vb{T}$ with eigenvalue $\lambda$:
\begin{equation} \label{eq:no_eigenspace}
\begin{split}
     \vb{T} \cdot (\vb{u} + \vb{v})^3 &= \vb{T} \cdot \vb{u}^3 + 3 \vb{T} \cdot \vb{u}^2 \vb{v} + 3 \vb{T} \cdot \vb{u} \vb{v}^2 + \vb{T} \cdot \vb{v}^3 \\
     &= \lambda (\vb{u} +\vb{v}) + 3 \vb{T} \cdot (\vb{u}^2 \vb{v} + \vb{u} \vb{v}^2),
\end{split}
\end{equation}
and the latter term in \cref{eq:no_eigenspace} does not generally vanish. This is reflective of the fact that the notion of eigenspaces is replaced by eigenvarieties (typically sets of discrete points) for higher-order symmetric tensors.

\subsubsection{Representing symmetric tensors: monomials and spherical harmonics.}

\begin{figure}[t]
    \centering
    \includegraphics[width=0.30\linewidth,valign=c]{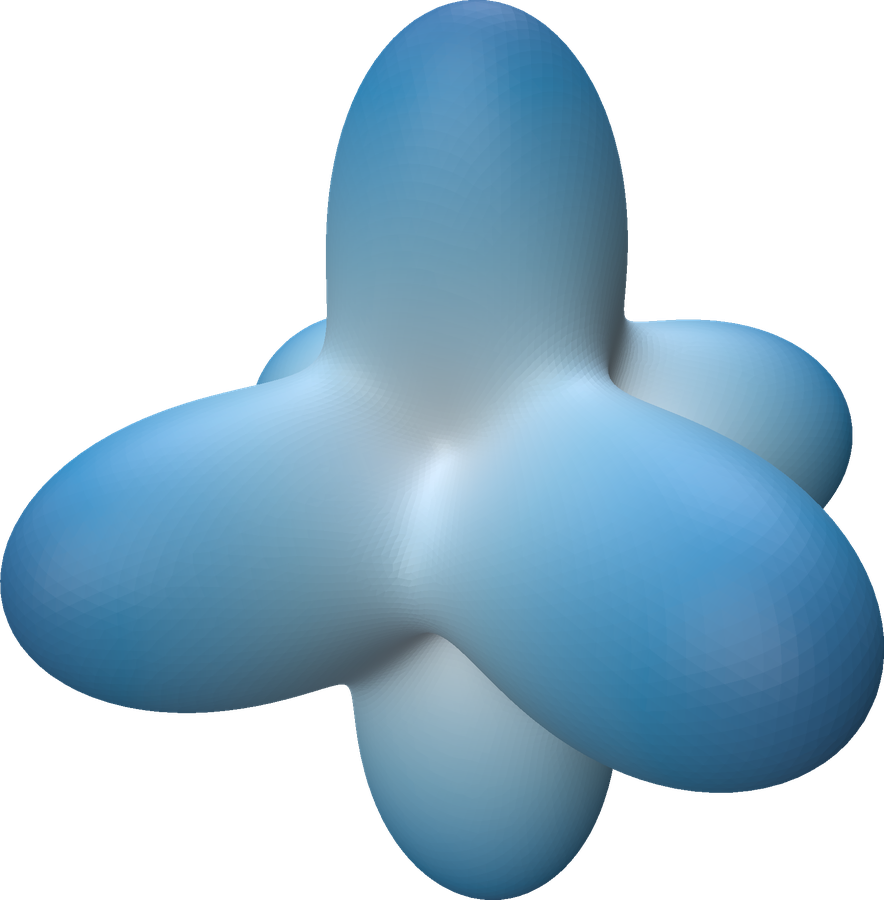} \hfill
    \includegraphics[width=0.30\linewidth,valign=c]{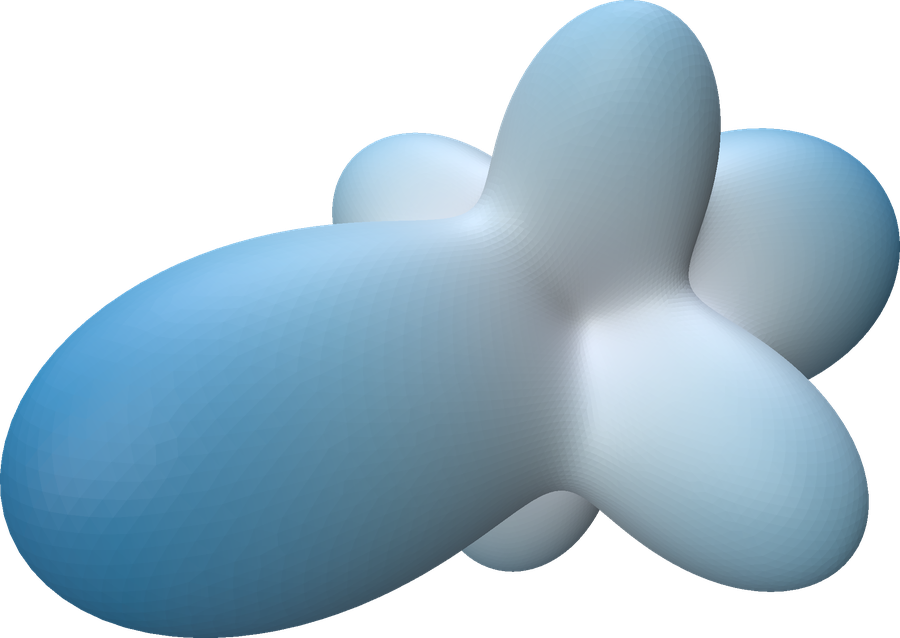} \hfill
    \includegraphics[width=0.26\linewidth,valign=c]{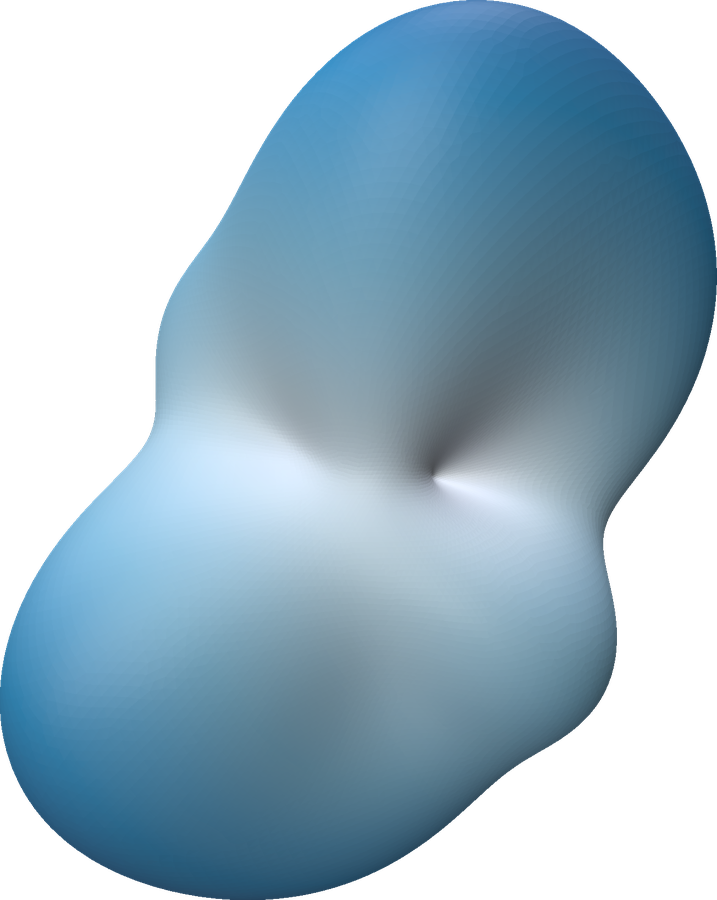}
    \caption{Visualization of tensor polynomials $p_{\vb{T}}(\vb{x})$. Left: an isotropic odeco tensor with equal eigenvalues. Center: an anisotropic odeco tensor with distinct eigenvalues. Right: a non-odeco tensor that does not decompose into orthogonal rank-1 terms.}
    \label{fig:odeco_polys}
\end{figure}

As done in \cite{Palmer20Algebraic}, we represent such symmetric tensors as spherical polynomials expressed in a basis of spherical harmonics. Note first that symmetric order-$d$ tensors $\vb{T}$ on $\bb{R}^n$ are determined fully by their values $\vb{T} \cdot \vb{x}^d$ for unit vectors $\vb{x} \in \mathbb{S}^{n-1}$. This holds by a generalized polarization identity and generalizes the correspondence between symmetric bilinear forms and quadratic forms (for $d=2$). In particular, if $\vb{x} = (x_1, \ldots, x_n)$, then: 
\begin{equation} \label{eq:tensor_poly}
	\begin{aligned}
		  \vb{T} \cdot \vb{x}^d &= \sum_{j_1,\dots,j_d=1}^n T_{j_1 \dots j_d} x_{j_1} \dots x_{j_d} \\
		&\hspace{-0.7cm}= \sum_{i_1+\dots+i_n=d} 
		\underbrace{{d \choose i_1,\dots,i_n} T_{\underbrace{\scriptstyle 1 \dots 1}_\text{$i_1$ times} \dots \underbrace{\scriptstyle n \dots n}_\text{$i_n$ times}}}_{u_{i_1 \dots i_n}} x_1^{i_1} \dots x_n^{i_n} \eqcolon p_{\vb{T}}(x_1,\dots,x_n).			
	\end{aligned}
\end{equation}
In the above, we see that a homogeneous degree $d$ polynomial $p_{\vb{T}}$ results, with coefficients denoted by $u_{i_1 \dots i_n}$. This expression takes into account the symmetries of the tensor coefficients, and shows the equivalence of such symmetric tensors to homogeneous degree-$d$ polynomials in $\bb{R}^n$. It also clearly shows the dimensionality of the space, which is equal to the number of multinomial coefficients $\binom{d}{i_1, \dots, i_n}$. Relevant to us are the case of order-4 symmetric tensors in 2D and 3D, which form 5- and 15-dimensional vector spaces, respectively. 
\Cref{fig:odeco_polys} shows graphs of such tensor polynomials restricted to $\bb{S}^2$ for odeco and non-odeco tensors.

Rather than use a basis of monomials as suggested by \Cref{eq:tensor_poly}, we can restrict $p_{\vb{T}}$ to $\bb{S}^{n-1}$ and use a basis of \emph{spherical harmonics}. Note that the restriction of the tensor polynomials to $\bb{S}^{n-1}$ also allows for a natural $L^2$ inner product (and thus metric) to be defined: 
\begin{equation}
	\langle p_{\vb{T}_1}, p_{\vb{T}_2} \rangle = \int_{\bb{S}^{n-1}} p_{\vb{T}_1}(\vb{x}) p_{\vb{T}_2}(\vb{x}) \, \dd\vb{x}.
\end{equation}
Let us concentrate on the $n=3$-dimensional case first.
\emph{Spherical harmonics} are a special set of functions on $\bb{S}^{n-1}$ that are orthonormal with respect to this $L^2$ inner product:
\begin{equation}
	Y_l^m(\vb{x}), \quad l = 0,1,\dots, \quad m=-l,\dots,l,
\end{equation}
where $l$ is the \emph{band} of the harmonic and $m$ its \emph{order}. The space of homogeneous degree 4 polynomials (restricted to $\bb{S}^2$) can be spanned by the spherical harmonics of bands $l=0,2,4$,
which we write compactly as
\begin{equation}
	p_{\vb{T}}(\vb{x}) = \sum_{l\in\{0,2,4\}} \sum_{m=-l}^l (q_{\vb{T}})_l^m Y_l^m(\vb{x}).
\end{equation}
Note that the number of terms in each band indeed sum to $1 + 5 + 9 = 15$, and the spherical harmonics form a basis. As they are orthonormal, the inner product and distance between tensors $\vb{T}_1$ and $\vb{T}_2$ can be calculated directly in terms of these coefficients: 
\begin{equation} \label{eq:tensor_distance}
	\langle p_{\vb{T}_1}, p_{\vb{T}_2} \rangle = \langle \vb{q}_{\vb{T}_1}, \vb{q}_{\vb{T}_2} \rangle \quad \text{and} \quad d(\vb{T}_1, \vb{T}_2) = \norm{\vb{q}_{\vb{T}_2} - \vb{q}_{\vb{T}_1}},
\end{equation}
where the $(q_{\vb{T}})_l^m$ have been gathered into vectors $\vb{q}_{\vb{T}} \in \bb{R}^{15}$. 

For the $n=2$ case, spherical harmonics have a 2D counterpart, sometimes called \emph{circular harmonics}, which correspond to the basis functions of a \emph{Fourier series}:
\begin{equation}	\underbrace{1}_{\text{band 0}}, \underbrace{\cos(2\theta), \sin(2\theta)}_{\text{band 2}}, \underbrace{\cos(4\theta), \sin(4\theta)}_{\text{band 4}}.
\end{equation}

\subsection{Odeco variety and band interpretations}

With the spaces of symmetric tensors parameterized, we must also be able to specify the subset of odeco tensors~\Cref{eq:odeco_defn}. In 3D, these form a variety defined by 27 quadratic relations on the monomial coefficients $u_{i_1,i_2,i_3}$ (Theorem 4 of \cite{boralevi_orthogonal_2017}):
\begin{equation} \label{eq:odeco_variety}
c_i(\vb{u}) \coloneq \vb{u}^\intercal A_i \vb{u} = 0,
\end{equation}
where coefficients have been gathered into a vector $\vb{u} \in \bb{R}^{15}$. These will be used to define an odeco constraint energy in \Cref{sec:odeco_relax}. The specific matrices $A_i$ may be found in various sources, e.g., the supplementary material of \cite{Palmer20Algebraic}.

\cite{Palmer20Algebraic} have also shown that the 0th and 2nd bands of spherical harmonics have interpretable meanings for odeco tensors. Band 0 encodes the tensor \emph{size}, since 
\begin{equation}
	(q_{\vb{T}})_0 = C_0 \sum_m \lambda_m,
\end{equation}
for a constant $C_0$. Band 2 encodes the tensor \emph{anisotropy}, since
\begin{equation} \label{eq:q2}
	\norm{(\vb{q}_{\vb{T}})_2}^2 = C_2 \sum_{m=1}^n (\lambda_{m+1} - \lambda_m)^2 \quad (m \in \bb{Z} \backslash n \bb{Z}),
\end{equation}
for a constant $C_2$. These relations and interpretations hold for both 2D and 3D odeco tensors. In particular, we note that a tensor is isotropic (eigenvalues all equal) if and only if $\norm{\vb{q}_2} = 0$, a fact used in \Cref{sec:dist_metrics} to define an angle distortion energy.

\subsubsection{Second-order part of odeco tensors.}
A very useful tool when manipulating odeco tensors is to look at its \emph{second-order part} $\vb{M}$,
defined as the second-order tensor (or $ n \times n $ matrix)
\begin{equation}
	M_{ij} = T_{ijkk},
\end{equation}
obtained by contracting two of its indices
\revision{(from here on, we use the Einstein summation convention)}.
Note that, thanks to the symmetry of $\vb{T}$, it does not matter on which two indices the contraction is done,
and $\vb{M}$ is a symmetric matrix.
Plugging in the odeco definition,~\cref{eq:odeco_defn}, yields
\begin{gather}
	M_{ij} = T_{ijkk} = \sum_{m=1}^n \lambda_m \left(\sum_{k=1}^n u_i^m u_j^m u_k^m u_k^m \right) = \sum_{m=1}^n \lambda_m u_i^m u_j^m, \\ \text{and} \quad
	\vb{M} = \sum_{m=1}^n \lambda_m (\vb{u}_m)^{\otimes2},
\end{gather}
where we used the fact that vectors $\vb{u}_m=\sum_i u^m_i \vb{e_i}$ have unit norm. 
In other words, the second-order part of a fourth-order odeco tensor
is a matrix that shares eigenvalues and eigenvectors
$\lambda_1\vb{u}_1, \dots, \lambda_n\vb{u}_n$ with the tensor.
This matrix is very useful since it allows us to scale the eigenvectors of odeco tensor $\vb{T}$.
First notice that taking matrix $\vb{M}$ to the power $p$ changes its eigenvalues accordingly:
\begin{equation}
	\vb{M}^p = \sum_{m=1}^n \lambda_m^p \vb{u}_m^{\otimes 2}.
\end{equation}
And contracting the tensor (once) with this matrix yields a fourth-order tensor with modified sizes
\begin{equation} \label{eq:tensor_modified_sizes}
	\vb{T}^{p+1} \triangleq \vb{T} \cdot \vb{M}^p = \qty( \sum_{m=1}^n \lambda_m^p \vb{u}_m^{\otimes 2} ) 
	\cdot \qty( \sum_{m=1}^n \lambda_m \vb{u}_m^{\otimes 4} )
	= \sum_{m=1}^n \lambda_m^{p+1} \vb{u}_m^{\otimes 4}.
\end{equation}
Integer power $p$ can be chosen arbitrarily and matrix $\vb{M}^p$ can be computed
using standard matrix algorithms;
in particular, $p=-1$ gives a unitary (or \emph{octahedral}) frame tensor $\vb{T}^0$
and $p=-2$ inverts the lengths of the frame vectors.
We demonstrate in~\cref{sec:odeco_integrability} how this tool enables us to properly
normalize the integrability energy.

\subsubsection{Surface/feature alignment, sizing, and isotropy.} \label{sec:alignment}
It is crucial to be able to impose alignment of a frame field to the surface or feature curves, as well as sizing of frame axes along these directions;
we explain how it is done in the odeco tensor setting. 

As noted in \cite{Palmer20Algebraic}, rotation of a spherical polynomial may be done by rotating its spherical harmonics coefficients in $\bb{R}^{15}$. These 15-dimensional rotation matrices are the \emph{Wigner D-matrices},
an essential tool of quantum physics for the study of angular momentum.
Consider the space of odeco frames $\vb{q}$ that are aligned to axis $\vb{\hat{e}}_z$,
and have a length of 1 in this direction.
This space can be parametrized by the affine transformation
\begin{equation} \label{eq:affine_transform}
	\vb{q} = \vb{q}_z + \vb{B}_z \vb{s},
\end{equation}
where $\vb{s} \in \bb{R}^5$ represents a 2D odeco tensor,
and $\vb{q}_z \in \bb{R}^{15}$ and $\vb{B}_z \in \bb{R}^{15 \times 5}$ are constants (see Section 5.1 of \cite{Palmer20Algebraic}). Having a length/sizing constraint $l$ different from 1 merely scales the vector $\vb{q}_z$ in the expression above.
Then, the space of odeco frames that are aligned to an arbitrary surface normal $\vb{n}$,
is obtained by applying any rotation matrix $\vb{R}_{\vb{n}} \in \bb{R}^{15 \times 15}$
that brings $\vb{\hat{e}}_z$ to $\vb{n}$:
\begin{equation} \label{eq:n_aligned_param}
	\vb{q} = \vb{R}_{\vb{n}} (\vb{q}_z + \vb{B}_z \vb{s}) = \vb{q}_{\vb{n}} + \vb{B}_{\vb{n}} \vb{s}.
\end{equation} 
\revision{For ease of implementation,}
instead of such an affine parametrization,
we are interested in finding an affine equation
\begin{equation} \label{eq:align-affine}
	\vb{A}_{\vb{n}} \vb{q} + \vb{b}_{\vb{n}} = \vb{0}
\end{equation}
that is satisfied by any $\vb{q}$ respecting the alignment constraint.
We propose a simple procedure that does not require figuring out $\vb{R}_{\vb{n}}$ nor $\vb{B}_z$
(or even its dimension),
and can be adapted to any kind of alignment constraint (e.g., feature edge or corner constraints).
First, one forms a matrix $\vb{Q}$ that contains as columns a batch of random tensors respecting the \emph{linear part}
of the alignment constraint; in this case, frames aligned to $\vb{n}$ but having zero length in that direction.
Let $\vb{N}$ have columns that form a basis of the null space of $\vb{Q}^\intercal$,
such that $\vb{Q}^\intercal \vb{N} = \vb{0}$;
we then have $\vb{A}_{\vb{n}} = \vb{N}^\intercal$.
Now let $\vb{q}_0$ be any tensor respecting the affine constraint;
then $\vb{b}_{\vb{n}} = - \vb{A}_{\vb{n}} \vb{q}_0$.
This procedure gives us the affine equation,~\cref{eq:align-affine}. 

\emph{Isotropy} of a tensor, i.e., all eigenvalues being equal, amounts to a linear constraint $(\vb{q}_{\vb{T}})_2 = \vb{0}$,
as shown by \cref{eq:q2}.
Given an $\vb{n}$-aligned tensor parametrized by \cref{eq:n_aligned_param},
isotropy in the plane orthogonal to $\vb{n}$ thus amounts to zeroing out these coefficients for the planar 2D odeco tensor $\vb{s}$, leaving an affine space
\begin{equation}
    \vb{q} = \vb{q}_{\vb{n}} + \vb{B}_{\vb{n}}^\rm{iso} \vb{s}^\rm{iso}
\end{equation}
of dimension 3, i.e., $\vb{B}_{\vb{n}}^\rm{iso} \in \bb{R}^{15\times3}$ and $\vb{s}^\rm{iso} \in \bb{R}^{3}$.
We denote the corresponding affine constraint equations \begin{equation} \label{eq:isotropy_constraint}
    \vb{A}_{\vb{n}}^\rm{iso} \vb{q} + \vb{b}_{\vb{n}}^\rm{iso} = \vb{0}.
\end{equation}

\subsubsection{Eigenvalue sensitivity for tensors.} \label{sec:eigensensitivity}

A last tool we add to our "tensor toolbox" is a result on eigenvalue sensitivity.
It answers the question: ``\emph{Given an odeco tensor, how do its eigenvectors change
when slightly perturbing its coefficients?}''
This tool will help us express integrability in terms of tensor coefficients,
as it can transform spatial derivatives on the frame vectors into spatial derivatives on the tensor.

This \emph{eigenvalue perturbation problem} is well-known for matrices,
but is easily generalized to higher-order tensors, provided they are odeco.
Consider an odeco tensor $\vb{T} = \sum_{m=1^n} \lambda_m \vb{u}_m^{\otimes 4}$, or, in index notation,
\begin{equation}
	T_{j_1 j_2 j_3 j_4} = \sum_{m=1}^n \lambda_m u^m_{j_1} u^m_{j_2} u^m_{j_3} u^m_{j_4}.
\end{equation}
Consider the contraction of $\vb{T}$ with one of its eigenvectors, $\vb{u}_k$:
\begin{equation}
	\begin{aligned}
		T_{j_1 j_2 j_3 j_4} u^k_{j_4} &= \sum_{m=1}^n \lambda_m u^m_{j_1} u^m_{j_2} u^m_{j_3} u^m_{j_4} u^k_{j_4} \\
		&= \sum_{m=1}^n \lambda_m u^m_{j_1} u^m_{j_2} u^m_{j_3} \delta_{mk} \\
		&= \lambda_k u^k_{j_1} u^k_{j_2} u^k_{j_3},
	\end{aligned}
\end{equation}
or, written compactly, $\vb{T}\cdot\vb{u}_k = \lambda_k \vb{u}_k^{\otimes 3}$.
Contracting again and following the same procedure we find that
$\vb{T}\cdot\vb{u}_k^{\otimes 2} = \lambda_k \vb{u}_k^{\otimes 2}$
and $\vb{T}\cdot\vb{u}_k^{\otimes 3} = \lambda_k \vb{u}_k$,
the latter being the definition of a tensor eigenvector.
Consider now a specific eigenpair $(\lambda,\vb{w})$,
with $\vb{w}$ having unit norm.
We wish to express the variation of the eigenpair $(\delta\lambda, \delta\vb{w})$
given some perturbation of the tensor $\delta\vb{T}$.
We write the remainder of the proof in compact notation for brevity,
but the same steps can be performed in index notation.
Since $\vb{w}$ has unit norm, it is orthogonal to its variation:
\begin{equation}
	(\vb{w} + \delta\vb{w}) \cdot (\vb{w} + \delta\vb{w}) = 1 \quad \Rightarrow \quad \vb{w}\cdot\delta\vb{w} = 0,
\end{equation}
neglecting the higher-order terms $(\delta\vb{w} \cdot \delta\vb{w})$.
Writing the eigenvector definition $\vb{T}\vb{w}^{\otimes 3} = \lambda\vb{w}$
for the new eigenpair,
\begin{equation}
	\begin{aligned}
		(\vb{T}+\delta\vb{T})(\vb{w}+\delta\vb{w})^{\otimes3} &= \lambda\vb{w} + \delta(\lambda\vb{w}), \\
		\cancel{\vb{T}\vb{w}^3} + 3\vb{T}\vb{w}^2\delta\vb{w} + \delta\vb{T}\vb{w}^3 &=
		\cancel{\lambda\vb{w}} + \delta(\lambda\vb{w}), &\,&\text{(eigenvector definition)} \\
		\cancel{3 \lambda\vb{w}^2 \delta\vb{w}} + \delta\vb{T}\vb{w}^3 &= \delta(\lambda\vb{w}),
		&\,&\text{($\vb{T}\vb{w}^2 = \lambda\vb{w}^2$,  $\vb{w}\cdot\delta\vb{w} = 0$)}.
	\end{aligned}
\end{equation}
We find the final eigenvalue sensitivity result: $\delta(\lambda\vb{w}) = \delta\vb{T}\vb{w}^3$;
the variation in a scaled eigenvector is obtained
by contracting the tensor perturbation three times with the unit eigenvector
(note that an eigenvector's sensitivity only depends on itself and not explicitely on the other eigenvectors).
From this result we find the partial derivatives
\begin{equation} \label{eq:eig_sensitivity}
	\pdv{(\lambda w_i)}{T_{j_1 j_2 j_3 j_4}} = w_{j_1} w_{j_2} w_{j_3} \delta_{ij_4}.
\end{equation}
\Cref{sec:odeco_integrability} will show how this result is put to use to derive an integrability expression.

\section{Method} \label{sec:method}

Our method optimizes for a symmetric tensor field that is  odeco and integrable nearly everywhere, with sparse violations naturally arising to achieve singularities. A novel integrability energy is first derived in \Cref{sec:odeco_integrability}, discretization details are provided in \Cref{sec:discretization}, and an ``odeco-ness'' energy and distortion energies are provided in \Cref{sec:odeco_relax,sec:dist_metrics}. Lastly, the explicit solver used is described in \Cref{sec:solver_gradients} and frame field recovery is described in \Cref{sec:frame_recovery}.

\subsection{Integrable odeco fields} \label{sec:odeco_integrability}

On a surface $\cal{M} \subset \bb{R}^3$
we define a three-dimensional orthogonal frame field $\vb{F}(\vb{x})$
that is aligned with the surface normal $\vb{n}(\vb{x})$.
At each non-singular point $\vb{x} \in \cal{M}$,
a local lifting of frame vectors $ \vb{u}(\vb{x'}), \vb{v}(\vb{x'}), \vb{n}(\vb{x'}) $
can be defined on a small neighborhood $\cal{N}$ of $\vb{x}$.
We assume the frame field respects the local injectivity property:
$\det\begin{pmatrix} \vb{u} & \vb{v} & \vb{n} \end{pmatrix} > 0$;
the frame being orthogonal, this means that none of the frame vectors can vanish.
As stated by~\cref{eq:intrinsic-curl}, the frame field is \emph{integrable}
if its frame vectors $\vb{u}$ and $\vb{v}$ have their curl tangent to the surface:
\begin{equation} \label{eq:curl-free-uv}
    (\curl\vb{u})\cdot\vb{n} = (\curl\vb{v})\cdot\vb{n} = 0.
\end{equation}
This section constructs an integrability criterion $h_{\vb{T}}(\vb{x})$,
only depending on the tensor field and its derivatives,
that is zero when the curl-free condition is satisfied.
Let $\vb{u} = \lambda\hat{\vb{u}}$ be an eigenvector of tensor $\vb{T}$.
We develop its curl:
\begin{equation} \label{eq:eigvec-curl}
    \begin{aligned}
        (\curl\vb{u})_j &= \epsilon_{jab} \, \pdv{u_b}{x_a} & \; &\text{($\epsilon$: Levi-Civita symbol)} \\
        &= \epsilon_{jab} \, \pdv{u_b}{T_{k_1k_2k_3k_4}} \, \pdv{T_{k_1k_2k_3k_4}}{x_a} & \; &\text{(chain rule)} \\
        &= \epsilon_{jab} \, \hat{u}_{k_1} \hat{u}_{k_2} \hat{u}_{k_3} \delta_{k_4b} \, \pdv{T_{k_1k_2k_3k_4}}{x_a} & \; &\text{(sensitivity \cref{eq:eig_sensitivity})} \\
        &= \hat{u}_{k_1} \hat{u}_{k_2} \hat{u}_{k_3} \, \epsilon_{jab} \, \pdv{T_{k_1k_2k_3b}}{x_a} & \; &\text{(sum on $k_4$)} \\
        &= \hat{u}_{k_1} \hat{u}_{k_2} \hat{u}_{k_3} \, (\curl \vb{T}_{k_1k_2k_3})_j & \; &\text{(curl definition)},
    \end{aligned}
\end{equation}
where $\vb{T}_{k_1k_2k_3} \in \bb{R}^3$ is the vector with components $(T_{k_1k_2k_3 1}, T_{k_1k_2k_3 2}, T_{k_1 k_2 k_3 3})$.
We wish to make the frame vectors $u_k$ disappear on the right-hand side
to obtain an expression that only depends on the tensor coefficients instead.
To do so we multiply the expression by the $i$-th component of $\vb{u}$
(note that $i$ and $j$ are free indices here):
\begin{equation}
    u_i (\curl\vb{u})_j = \lambda_u \hat{u}_{k_1} \hat{u}_{k_2} \hat{u}_{k_3} \hat{u}_i \, (\curl \vb{T}_{k_1k_2k_3})_j.
\end{equation}
Summing this expression for each of the frame vectors $\vb{u}, \vb{v}, \vb{n}$ gives
\begin{equation}
    u_i(\curl{\vb{u}})_j + v_i(\curl{\vb{v}})_j + n_i(\curl{\vb{n}})_j = T_{k_1k_2k_3i}(\curl{\vb{T}_{k_1k_2k_3}})_j.
\end{equation}
Both sides of this identity can be seen as a 3-by-3 matrix.
We dot-product them with the surface normal $\vb{n}$, which gives a scalar equation for each $i$:
\begin{multline} \label{eq:curl_expression}
    u_i(\curl{\vb{u}})\cdot\vb{n} + v_i(\curl{\vb{v}})\cdot\vb{n} + n_i\overbrace{(\curl{\vb{n}})\cdot\vb{n}}^{=0} \\= T_{k_1k_2k_3i}(\curl{\vb{T}_{k_1k_2k_3}})\cdot\vb{n}.
\end{multline}
The third term is zero as the curl of the surface normal is always tangent to the surface.
Taking the $L^2$ vector norm on both sides we finally obtain
\begin{multline}
    \lambda_u^2 \qty((\curl{\vb{u}})\cdot{\vb{n}})^2 + \lambda_v^2 \qty((\curl{\vb{v}})\cdot{\vb{n}})^2 \\ = 
    \qty( \sum_{k_1,k_2,k_3=1}^3 \vb{T}_{k_1k_2k_3}  (\curl{\vb{T}_{k_1k_2k_3}}) \cdot \vb{n} )^2.
\end{multline}
This expression is purely in terms of tensor entries and may be considered for general symmetric tensors, even if they are not odeco. 

In \cite{Couplet26Size} (Figs. 10 and 11), it was shown that normalization of this kind of base expression, by a reciprocal squared power of eigenvector lengths, is required to have an energy that is independent of mesh resolution.
\revision{
Around singularities of index $k/4$, frame lengths grow at a rate of
$r^{-k/4}$, where $r$ is the radial distance from the singularity.
This means that index $+1/4$ (valence 3) singularities blow up, and index $-1/4$ (valence 5) vanish.
The normalization does not prevent either from appearing,
but introduces a fixed discretization error that does not favor either singularity type.
}
\revision{Empirically, the normalization also acts as a barrier preventing frame lengths to become negative.}
For the normalization in this setting: take \cref{eq:curl_expression}, multiply both sides by
$\vb{M}^{-2} = \sum_m \lambda_m^{-2} \vb{\hat{u}}_m^{\otimes2}$ and we get
\begin{equation}
    \frac{u_i}{\lambda_u^2}(\curl{\vb{u}})\cdot\vb{n} + \frac{v_i}{\lambda_v^2}(\curl{\vb{v}})\cdot\vb{n} = T^{-1}_{k_1k_2k_3i}(\curl{\vb{T}_{k_1k_2k_3}})\cdot\vb{n}.
\end{equation}
Taking the $L^2$ vector norm again gives
\begin{multline}
    \qty(\frac{(\curl{\vb{u}})\cdot{\vb{n}}}{\lambda_u})^2 + \qty(\frac{(\curl{\vb{v}})\cdot{\vb{n}}}{\lambda_v})^2 \\ = 
    \qty( \sum_{k_1,k_2,k_3=1}^3 \vb{T}^{-1}_{k_1k_2k_3}  (\curl{\vb{T}_{k_1k_2k_3}}) \cdot \vb{n} )^2\eqcolon h_{\vb{T}}.
\end{multline}
This finalizes our expression for our normalized integrability energy $h_{\vb{T}}$ used within our optimization. 
\revision{}

\subsection{Discretization and Optimization}

We formulate the search for an integrable frame field as an optimization problem
striving to minimize the integral of the integrability measure $h_{\vb{T}}$
over the domain $\Omega$,
\begin{equation}
    H[\vb{T}] = \int_\Omega h_{\vb{T}}(\vb{x}) \, \dd\vb{x},
\end{equation}
which we call the \emph{curl energy}.
We start by explaining how this integral is computed in practice,
then we present the optimization approach.

\subsubsection{Discretization and integration.} \label{sec:discretization}

The computations are supported by a standard triangle mesh in 2D.
At each mesh vertex we store the tensors as their spherical harmonics coefficients $\vb{q} \in \bb{R}^{15}$, with a normal alignment condition enforced at each vertex. 
These coefficients are interpolated throughout the domain by a continuous function $\vb{q}(\vb{x})$
that is piecewise linear on the mesh elements. 
Consider a triangle $\cal{T}$ with vertex positions $\vb{x}^1,\vb{x}^2,\vb{x}^3$,
vertex values $\vb{q}^1, \vb{q}^2, \vb{q}^3$
and corresponding linear shape functions $\psi^1(\vb{x}), \psi^2(\vb{x}), \psi^3(\vb{x})$
(such that $\psi^i(\vb{x}^j) = \delta_{ij}$).
The interpolant over the triangle is defined by
\begin{equation}
    \vb{q}(\vb{x}) = \sum_{i=1}^3 \vb{q}^i \, \psi^i(\vb{x}).
\end{equation}
Note while vertex tensors $\vb{q}^i$ are constrained to be normal-aligned and motivated to be odeco, these constraints/energies are not imposed on the interpolated tensors interior to triangles.
Computing the tensor curl $h_{\vb{T}}$ requires evaluating the spatial derivatives of $\vb{q}(\vb{x})$;
this is done by taking the gradient on both sides:
\begin{equation}
    \grad\vb{q}(\vb{x}) = \sum_{i=1}^3 \vb{q}^i \grad \psi^i(\vb{x}).
\end{equation}
Note that this gradient is constant per element since the shape fuctions $\psi^i$ are linear.
In order to numerically integrate a function $f(\vb{q})$ over $\cal{T}$
(for example, the integrability measure $h_{\vb{T}}$),
we apply a 3-point Gaussian quadrature rule
\begin{equation}
    \int_\cal{T} f(\vb{q}(\vb{x})) \, \dd\vb{x} \approx
    \sum_{k=1}^3 w_k \, f(\vb{q}(\vb{x}(\vb{\xi}_k))) \, J_\cal{T},
\end{equation}
where $\vb{\xi}_i$ are the quadrature points on a reference triangle,
$w_i$ the corresponding quadrature weights,
and $J_\cal{T}$ the determinant of the Jacobian of the transformation $\vb{x}(\vb{\xi})$
mapping the reference triangle to $\cal{T}$.
This 3-point quadrature rule has a degree of precision of 2,
meaning that it is exact for quadratic polynomials.
The integrability measure $h_{\vb{T}}$ is not a polynomial
but we found that this degree of precision is accurate enough for our purposes.
The integral of $f$ over the whole domain is then obtained by summing the integral over the triangles
\begin{equation}
    \int_\cal{S} f(\vb{q}(\vb{x})) \, \dd\vb{x} = \sum_i \int_{\cal{T}_i} f(\vb{q}(\vb{x})) \, \dd\vb{x}.
\end{equation}
This calculation is embarrassingly parallel,
as the sum over the triangles can be distributed over a large number of CPUs.

\revision{
Note that evaluating the integrability measure $h_{\vb{T}}$ at the quadrature points $\xi_i$
requires an estimate of the surface normal $\vb{n}$.
For this purpose we simply use the triangle normal $\vb{n}_\cal{T}$.
}

\begin{figure}[htbp]
    \centering
    \begin{tabular}{@{}m{1.5em}@{\hspace{0.5em}}c@{\hspace{0.5em}}c@{}}
        & Minimizing area distortion & Minimizing angle distortion \\[1em]
        & \scriptsize$-1$\,\includegraphics[height=1em, valign=m]{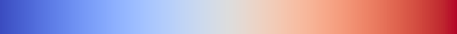}\,$1$ & \scriptsize$0$\,\includegraphics[height=1em, valign=m]{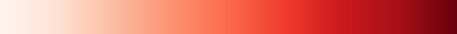}\,$1$ \\[1em]
        \rotatebox[origin=c]{90}{Mambo M1} &
        \includegraphics[width=0.45\linewidth, valign=m]{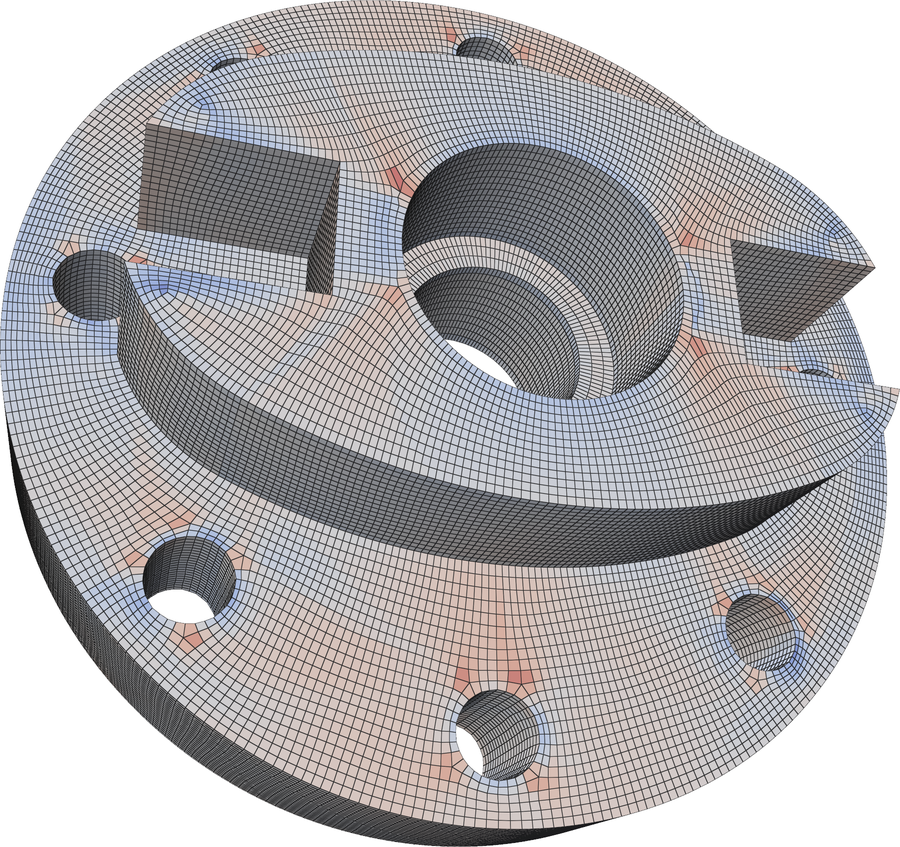} &
        \includegraphics[width=0.45\linewidth, valign=m]{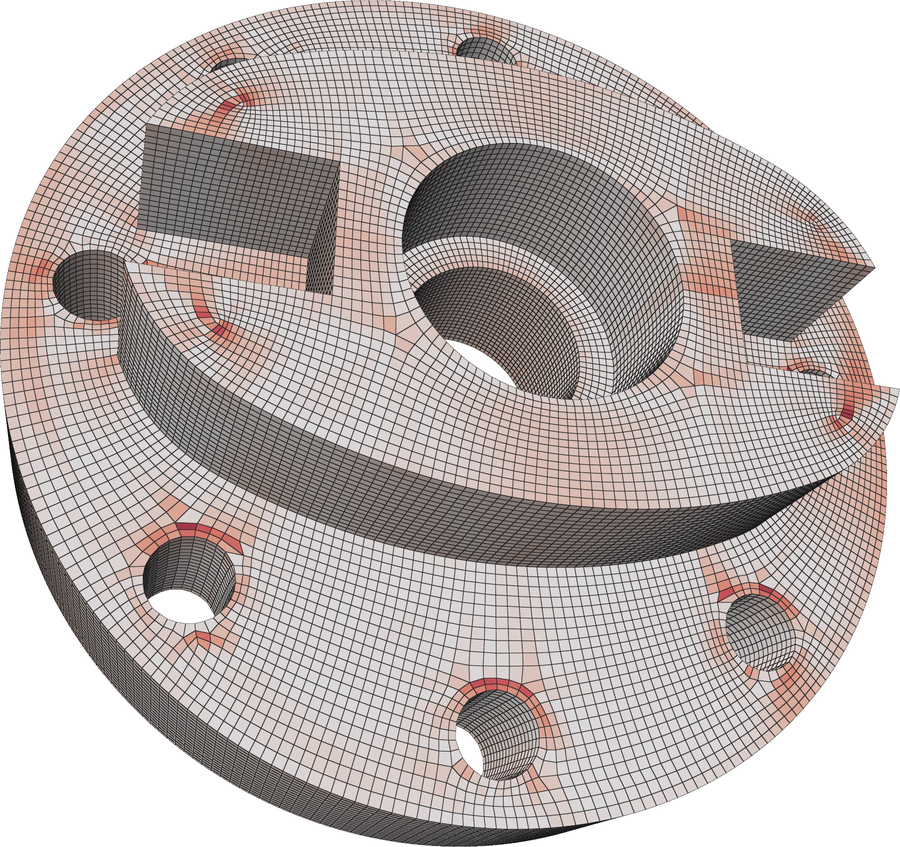} \\ \addlinespace[2em]
        \rotatebox[origin=c]{90}{Mambo M2} &
        \includegraphics[width=0.45\linewidth, valign=m]{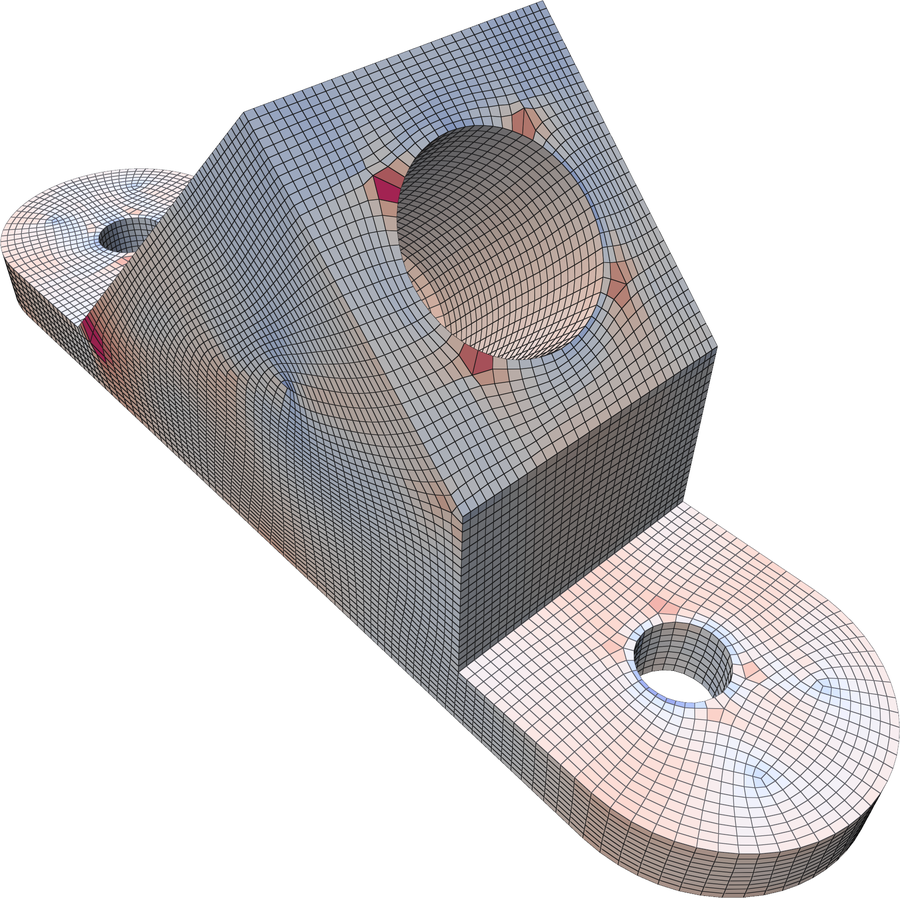} &
        \includegraphics[width=0.45\linewidth, valign=m]{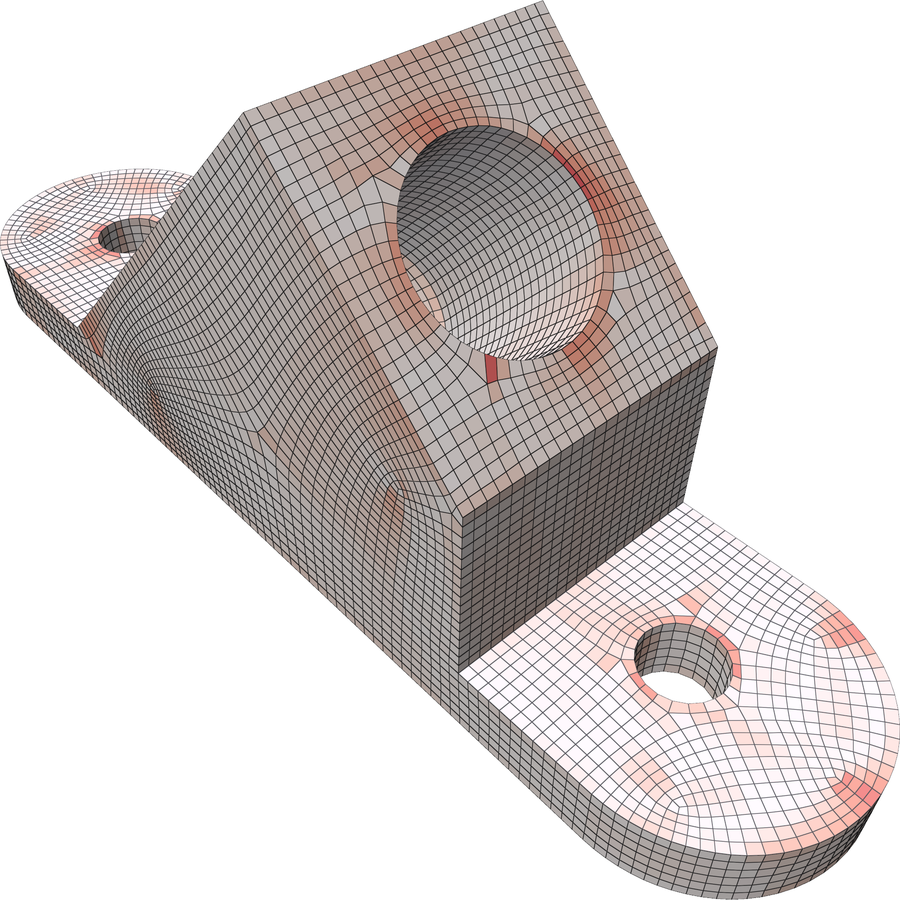} \\ \addlinespace[2em]
        \rotatebox[origin=c]{90}{Fertility} &
        \includegraphics[width=0.45\linewidth, valign=m]{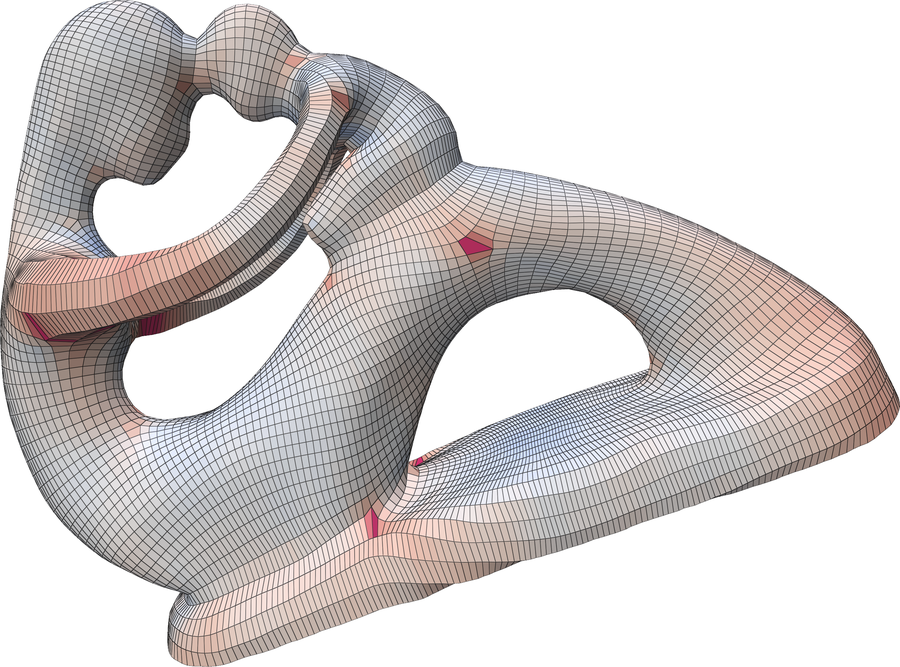} &
        \includegraphics[width=0.45\linewidth, valign=m]{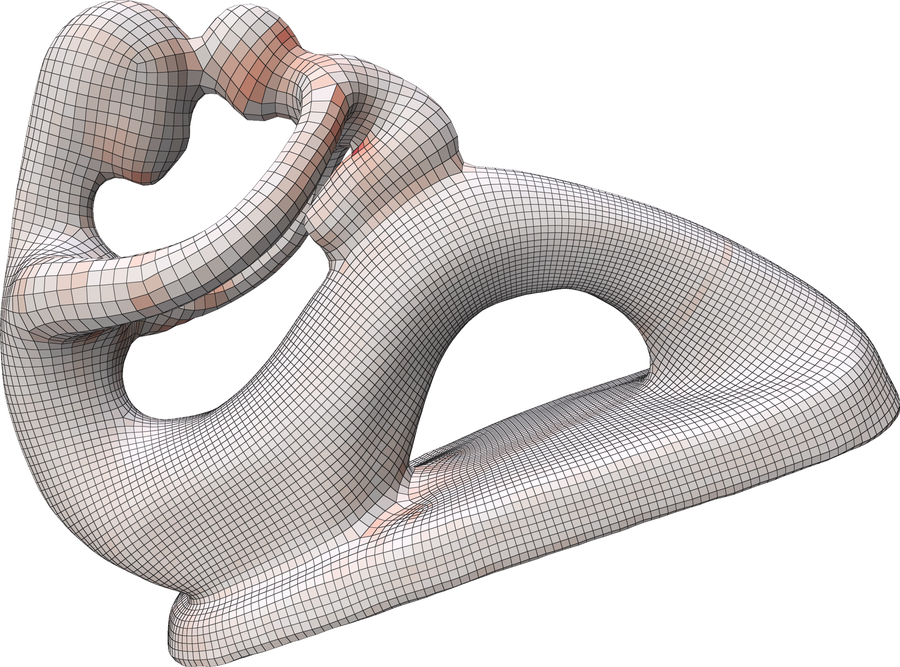} \\ \addlinespace[2em]
        \rotatebox[origin=c]{90}{Botijo} &
        \includegraphics[width=0.40\linewidth, valign=m]{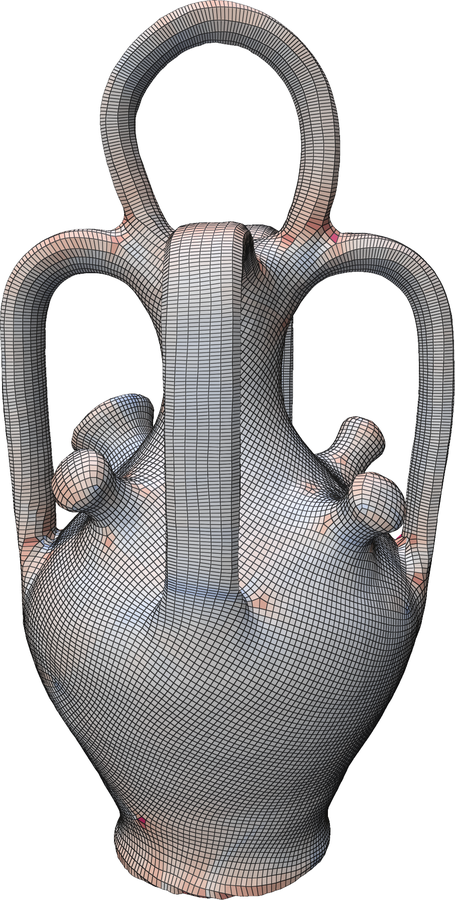} &
        \includegraphics[width=0.40\linewidth, valign=m]{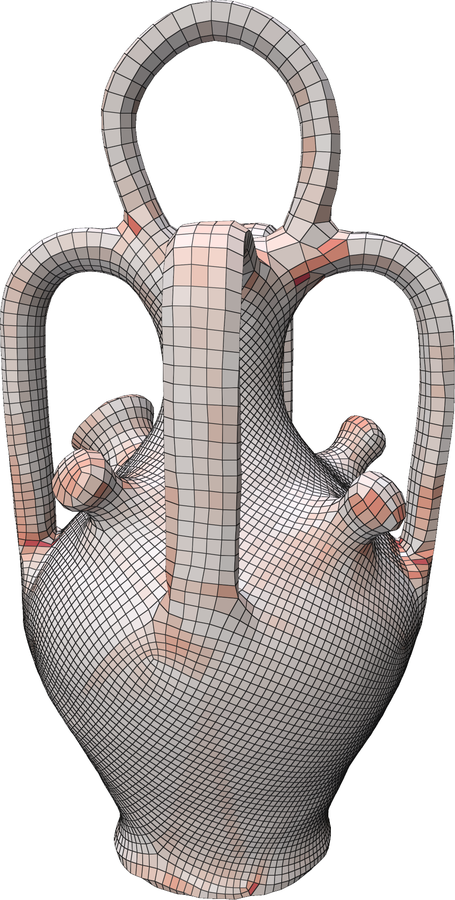} \\
    \end{tabular}
    \caption{Quad mesh results with area- and angle-distortion metrics on select CAD models (feature alignment, but no sizing constraints) and smooth models (no features) using different distortion energies. Note the higher anisotropy in the area-preserving mode and the differing singularity placements between the different modes of the framework. } 
    \label{fig:area-angle-comparison}
\end{figure}

\subsubsection{Relaxation of the odeco constraints.} \label{sec:odeco_relax}

At this point, it might be tempting to simply minimize $H[\vb{T}]$
under the constraint that the field is everywhere odeco.
However, this would fail to introduce new singularities
as it is often energetically cheaper to distribute the curl
on the domain rather than to introduce the fixed discretization cost of singularities
\revision{(this fixed cost comes from the fact that a piecewise linear representation
cannot capture the frame vectors blowing up when nearing singularities)}.
To address this, we relax the odeconess constraint and allow
the solver to introduce non-odeco terms that reduce the integrability cost
in the neighborhood of singularities.
A natural choice would be to introduce an \emph{odeco penalty}
\begin{equation}
	C[\vb{T}] \triangleq \sum_i \int_\Omega c_i^2(\vb{T}(\vb{x})) \, \dd\vb{x},
\end{equation}
where $c_i(\cdot)$ are the quadrics defining the odeco variety referred to in \cref{eq:odeco_variety};
this functional is thus a 4th-order polynomial in the tensor coefficients. 

In practice, we introduce a \emph{normalized odeco penalty} that applies an $L^2$ normalization before application of squared constraint equations:
\begin{equation}
	\hat{C}[\vb{T}] \triangleq \sum_i \int_\Omega c_i^2\left(\frac{\vb{T}}{\norm{\vb{T}}}\right) \, \dd\vb{x},
\end{equation}
Empirically, this prevents the shrinking of the tensors, as the standard odeco penalty $C[\vb{T}]$ can be minimized by setting $T=0$.

\revision{Note that most results on tensor-based integrability
assume a tensor field that is odeco everywhere.
The relaxation breaks this property.
Our core assumption is that notions of integrability and distortion
still make sense when the field is not perfectly odeco,
and this is validated by experiments.
}

\subsubsection{Distortion metrics.} \label{sec:dist_metrics}

Lastly, we optionally include distortion metrics that aim to preserve the area of quad elements, or aim to minimize the angle distortion, promoting isotropic (conformal) quad elements. Recall that $\det\,\vb{M} = \Pi_m \lambda_m$, which is the inverse of frame volume or area (as the eigenvalues represent gradient norms).
\begin{align}
    \label{eq:area_dist} \Phi_\rm{area}\qty[\vb{T}] &= \int_\Omega \log^2\left(A_0\,\revision{\det\,\vb{M}}\right) \, \dd\vb{x}, \\
    \label{eq:angle_dist} \Phi_\rm{angle}\qty[\vb{T}] &= \int_\Omega \norm{\vb{b}_{\vb{n}}^\rm{iso} - \vb{A}_{\vb{n}}^\rm{iso} \vb{q}}^2 \, \dd\vb{x}.
\end{align}
Above, $A_0$ denotes the target area of quad elements, and $\vb{n}$ is the normal of the triangle we're integrating over. Recall that $\vb{A}_{\vb{n}}^\rm{iso}$ and $\vb{b}_{\vb{n}}^\rm{iso}$ were defined in \cref{eq:isotropy_constraint}. Rows of $\vb{A}_{\vb{n}}^\rm{iso}$ are made orthonormal so that this objective measures actual distance to the affine space.

\subsubsection{Full objective and initialization}
The combination of the normalized integrability and odeco energies and the optional distortion energies leads us to our full objective function:
\begin{equation} \label{eq:E_kappa}
	E[\vb{T}] \coloneq H[\vb{T}] +  \kappa_{\text{odeco}}\hat{C}[\vb{T}] + \kappa_{\text{area}} \Phi_{\text{area}}[\vb{T}] + \kappa_{\text{angle}} \Phi_{\text{angle}}[\vb{T}].
\end{equation}
The $\kappa$ weights used in our results vary depending on the desired amount of anisotropy in the end mesh. The exact settings are described at the start of \Cref{sec:results}. 

Note that if $L$ is a characteristic size for the domain $\Omega$,
then the energies $\hat{C}$, $\Phi_\rm{area}$ and $\Phi_\rm{angle}$ scale with $L^2$
whereas $H$ does not scale.
This can be problematic when working with models at different scales.
To achieve scale independence, we divide the integrands by the local triangle area $J_\cal{T}$.
This ensures that the different energies remain adequately weighted regardless of scale.

As an initial solution, we solve for a smooth octahedral (i.e., odeco and unit norm) field.
Smoothness is achieved by a Dirichlet energy $\Phi_\rm{smooth}$ defined as in \cite{Palmer20Algebraic}:
\begin{equation}
    \Phi_\rm{smooth}[\vb{T}] = \sum_j \frac{1}{2} \int_\Omega \norm{\grad q_j}^2 \, \dd\vb{x}.
\end{equation}
The octahedral variety is cut out from the odeco variety by an affine space $\vb{A}^\rm{octa}\vb{q} + \vb{b}^\rm{octa} = \vb{0}$,
which can be computed in the same way as presented in \cref{sec:alignment}.
The search for an initial smooth octahedral field can thus be formulated by minimizing
\begin{equation} \label{eq:E_init}
    E_\rm{init}[\vb{T}] \coloneq \Phi_\rm{smooth}[\vb{T}] + \kappa_\rm{odeco} \hat{C}[\vb{T}] \quad \text{s.t.} \quad \vb{A}^\rm{octa}\vb{q} + \vb{b}^\rm{octa} = \vb{0}.
\end{equation}
For this optimization we consistently set $\kappa_\rm{odeco}=1$.
Note that the initial solution may violate the sizing constraints we set in the actual optimization;
in that case we first project the tensors onto their corresponding affine constraint space.

\subsubsection{Boundary alignment.}
As seen in~\cref{sec:alignment},
any alignment or sizing constraint on an odeco tensor can be written as an affine expression at each node
\begin{equation} \label{eq:boundary-align}
    \vb{A} \vb{q} + \vb{b} = \vb{0},
\end{equation}
with $\vb{A} \in \bb{R}^{10\times 15}$ and $\vb{b} \in \bb{R}^{10}$.
As a reminder, we determine this linear subspace by randomly generating a batch of (odeco) tensors
respecting the constraint we want to impose,
which allows us to compute $\vb{A}$ and $\vb{b}$. The $5$-dimensional solution space reflect the 5 degrees of freedom afforded by the 2D tensor subspace orthogonal to the fixed frame along the alignment direction.

Such constraints are imposed at nearly every vertex of the mesh. On general surface points, we impose alignment to the surface normal, with sizing of 1. On feature curve points, alignment tangent to the curve is enforced, with user-specified sizing in that direction. Note that the normal direction alignment is not imposed and allowed to be free, as these features are usually sharp edges with ambiguous normal directions. On ``corners'' where multiple feature curves meet, no alignment conditions are imposed due to the even greater normal (and feature tangent) ambiguity.

\subsubsection{Solver and gradient computation.} \label{sec:solver_gradients}

The optimization problem we pose (minimizing \cref{eq:E_kappa} subject to~\cref{eq:boundary-align})
is a nonlinear and nonconvex problem with linear constraints.
We address it using a quasi-Newton method, namely the limited-memory BFGS algorithm,
which we modify to impose the linear constraints at each update.
Let $\vb{q}$ be a tensor at a node that respects its alignment constraint $\vb{A}\vb{q} + \vb{b} = \vb{0}$,
and let $\vb{g} \in \bb{R}^N$ be the gradient of the objective function with respect to $\vb{q}$.
To ensure that any L-BFGS update does not violate the constraint,
we project $\vb{g}$ onto the space normal to the rows of $\vb{A}$, effectively ensuring $\vb{A}\vb{g} = \vb{0}$.
Recalling that $\vb{A}$ was constructed to have orthonormal rows,
this projection is simply done by iteratively subtracting from $\vb{g}$ its component along row $\vb{a}_i$:
\begin{equation}
    \text{for $i=1,\dots,M$:} \quad \vb{g} := \vb{g} - (\vb{g}\cdot\vb{a}_i) \vb{a}_i.
\end{equation}
We use the L-BFGS implementation of ALGLIB \cite{alglib}.
Analytical gradients are computed via automatic differentiation, using the TinyAD library \cite{schmidt2022tinyad}.

\subsubsection{Frame field recovery.} \label{sec:frame_recovery}
The minimization of~\eqref{eq:E_kappa} provides
a field of tensors that is odeco everywhere except
in the vicinity of singularities. For the following mesh generation step, we recover a frame on every triangle face by projecting the linearly-interpolated tensors onto the odeco variety
using the semidefinite projector described in \cite{Palmer20Algebraic}. Let us denote the recovered field on each face $t$: 
\begin{equation}
    G_t \coloneq[\nabla u | \nabla v]\in \mathbb{R}^{3\times 2}, 
\end{equation} 
where $u,v$ refer to putative components of a seamless parameterization.

\subsection{Mesh generation} \label{sec:mesh_gen}
There is a one-to-one correspondence between quad meshes and \emph{integer-grid maps} (IGM) \cite{Bommes2013IGM}, which are discrete realizations of the integer seamless maps described in \Cref{sec:seamless}. Consequently, in our context the goal of the mesh generation stage consists of determining an IGM $\phi_{IGM}$, which is as similar as possible to the integrable odeco field $G$, i.e., a piecewise linear map $\phi_{IGM}$ minimizing
\begin{equation}
\label{eqn:IGM}
	E_{IGM} = \int_\Omega \norm{J_{\phi_{IGM}} - G}^2_2 \; dA
\end{equation}
subject to IGM constraints, including local injectivity, and integer-quantization of transition functions, singularities, and feature curves (cf.~\cite{Bommes2013IGM}). Please note that up to discretization errors, the integrable odeco field $G$ already coincides to a seamless map $\phi_S$ such that the main deviation to $\phi_{IGM}$ is induced by additional integer-quantization constraints, not being considered in the integrable odeco field optimization. 
We employ an established approach, conceptually identical to \cite{Lyon2019FreeBoundaries} Figure 4, and the robust pipeline described in \cite{Capouellez2025Feature} Section 8. The key idea is to decompose the challenging IGM generation task into a series of independent sub-steps, for which robust and fast algorithms are available.
In a nutshell our mesh generation algorithm consists of (i) generating a locally injective seamless map $\phi_{S}$ closely resembling the integrable odeco field $G$ by optimizing a QP induced by Eqn.(\ref{eqn:IGM}) in conjunction with the linearized local injectivity constraints of \cite{Bommes2013IGM}, (ii) resolving numerical imprecisions of the seamless map constraints with \cite{Mandad2019ECS}, (iii) generating a T-mesh partitioning by tracing the motorcycle complex \cite{Campen2015QGP}, (iv) integer-quantization of T-mesh arcs \cite{heistermann2023MinDevFlow}, (v) T-mesh per patch parametrization to the integer-quantized rectangles determined in the previous step resulting in an initial IGM $\phi_{IGM}$, (vi) global relaxation of $\phi_{IGM}$ by minimizing the symmetric Dirichlet distortion energy \cite{Smith2015Bijective} w.r.t.~deviation from $\phi_S$, while respecting the quantization constraints, (vii) extracting the quad mesh induced by $\phi_{IGM}$ with the robust algorithm of \cite{Ebke2013QEX}.

\revision{
\subsection{Overview of pipeline and capabilities} \label{sec:pipeline}
Putting everything together, our end-to-end meshing pipeline proceeds in the following main steps:
\begin{enumerate}[itemsep=1ex]
    \item 
        Compute a smooth \emph{octahedral} (i.e., unit odeco) field $\vb{T}^\rm{smooth}$
        with alignment constraints, \cref{eq:E_init}.
    \item 
        Using $\vb{T}^\rm{smooth}$ as initialization,
        compute an integrable \emph{odeco} field $\vb{T}^\rm{integ}$ with alignment and sizing constraints
        and optional distortion objective, \cref{eq:E_kappa}.
    \item 
        Recover frame field $G$ by linearly interpolating
        odeco tensors to every triangle face,
        and projecting onto the odeco variety using the semidefinite projector of \cite{Palmer20Algebraic} (\cref{sec:frame_recovery}).
    \item Compute integer-grid map and extract quad mesh (\cref{sec:mesh_gen}).
\end{enumerate}
}

\revision{
We summarize the constraints that can be prescribed by our algorithm:
\begin{description}[itemsep=1ex]
    \item[Alignment:] can be set strongly or weakly through linear constraints
    on the tensor coefficients (\cref{sec:alignment}).
    The optimization empirically aligns to principal curvature directions,
    and the user can enforce stricter curvature alignment constraints if desired.
    \item[Sizing:] can be prescribed in two ways:
    \begin{itemize}
        \item \emph{impose size together with a direction},
        either strongly or weakly; this is done by augmenting the alignment constraint,
        making it affine instead of linear.
        \item \emph{impose local area} in a weak sense through the area distortion energy,
        \cref{eq:area_dist}.
    \end{itemize}
    \item[Anisotropy:] can be controlled in a weak sense through the angle distortion energy,
    \cref{eq:area_dist}.
\end{description}
Future work will explore user prescription of singularity positions.
}

\section{Results} \label{sec:results}

We validate our method on two subsets: the 9 ``Medium'' and 1 ``Simple'' feature-rich CAD models from
the MAMBO dataset~\cite{mambo_dataset}, and 4 smooth models from the dataset of~\cite{Myles:2014:RFG}, sans feature curves. MAMBO is split into three categories of increasing difficulty: Basic, Simple, and Medium, so our choice reflects the most challenging set of models. All meshes obtained in our results and comparisons below are included as part of the supplementary materials with our submission. If this submission is accepted, we will release an open-source version of the code.

Computations were done on an Apple M3 Pro
with 12 threads parallelized by OpenMP.
We stop the L-BFGS iterations when the relative objective improvement is $< 10^{-5}$.
We used the following sets of weighting coefficients
for the optimization unless otherwise stated:
\begin{itemize}
    \item for \emph{area} distortion minimization on \emph{CAD models}
    we set $\kappa_\rm{odeco} = 10$ and $\kappa_\rm{area} = 0.1$ (and $\kappa_\rm{angle} = 0$),
    \item for \emph{area} distortion minimization on \emph{smooth models}
    we set $\kappa_\rm{odeco} = 10$, $\kappa_\rm{area} = 0.1$,
    and $\kappa_\rm{angle} = 0.0001$;
    the addition of a small amount of $\Phi_\rm{angle}$ prevents the solver from completely collapsing frames along one dimension,
    \item for \emph{angle} distortion minimization
    we set $\kappa_\rm{odeco} = 1$ and $\kappa_\rm{angle} = 0.01$ (and $\kappa_\rm{area} = 0$),
    \item for sizing constraints only (no distortion energies)
    we set $\kappa_\rm{odeco}=1$.
\end{itemize}
Empirically, we found that the higher degree of anisotropy that is typical in area-preserving schemes required a higher $\kappa_\rm{odeco}=10$ for best performance.

\subsection{Demonstration of Distortion Energies} \label{sec:distortion_demo}

\begin{figure}[thbp]
    \centering
    \begin{tabular}{@{}m{1.5em}@{\hspace{0.5em}}c@{\hspace{0.5em}}c@{}}
        & Minimizing area distortion & Minimizing angle distortion \\[0.5em]
        & \scriptsize$-1$\,\includegraphics[height=1em, valign=m]{figures/colorbar-coolwarm.png}\,$1$ & \scriptsize$0$\,\includegraphics[height=1em, valign=m]{figures/colorbar-reds.png}\,$1$ \\[0.5em]
        \rotatebox[origin=c]{90}{Retinal} &
        \includegraphics[width=0.45\linewidth, valign=m]{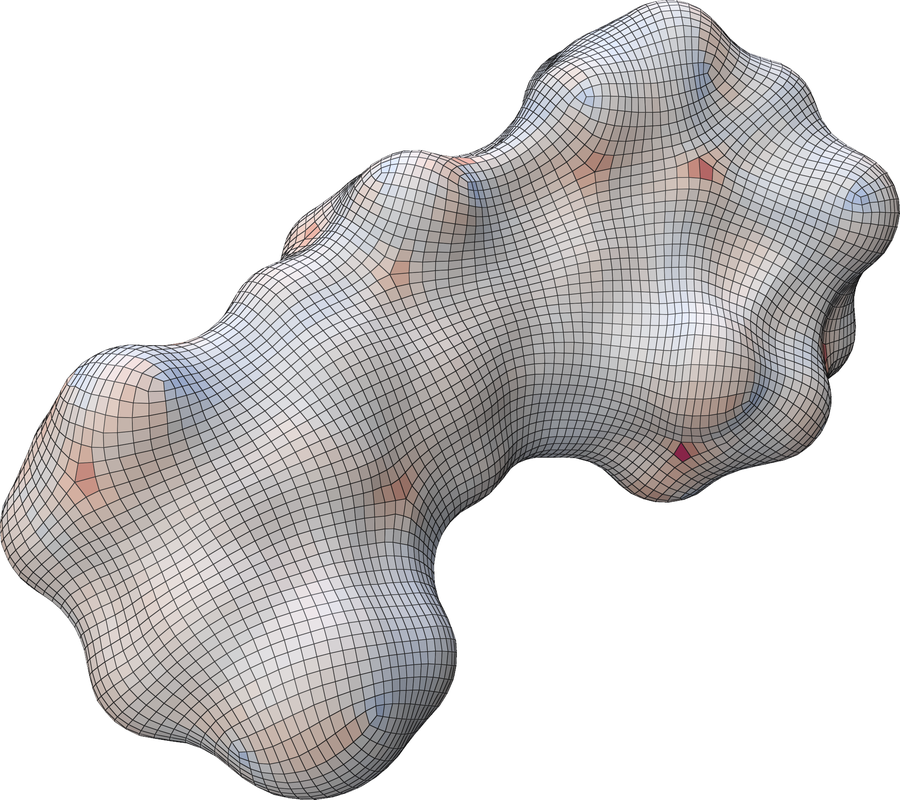} &
        \includegraphics[width=0.45\linewidth, valign=m]{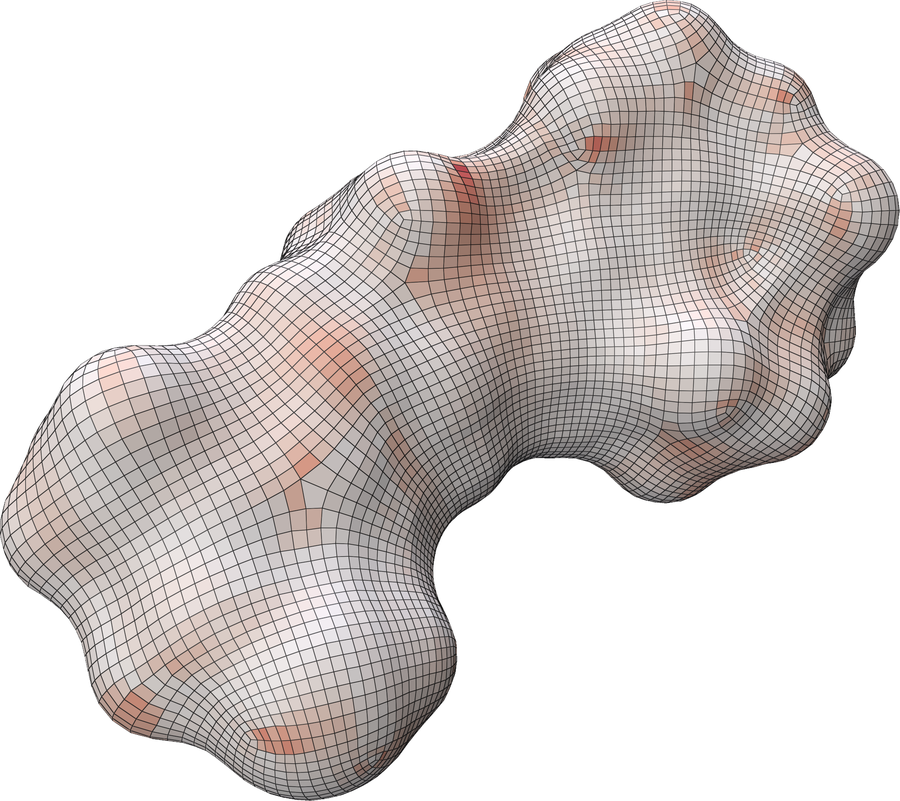} \\ \addlinespace[1em]
        \rotatebox[origin=c]{90}{Genus3} &
        \includegraphics[width=0.45\linewidth, valign=m]{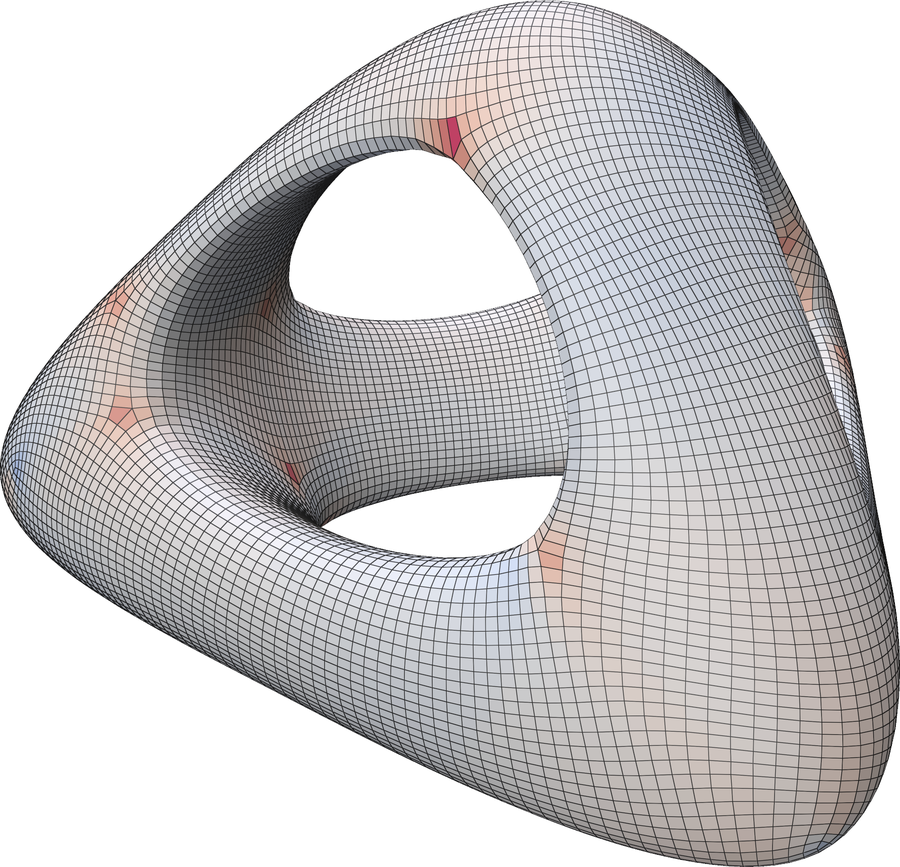} &
        \includegraphics[width=0.45\linewidth, valign=m]{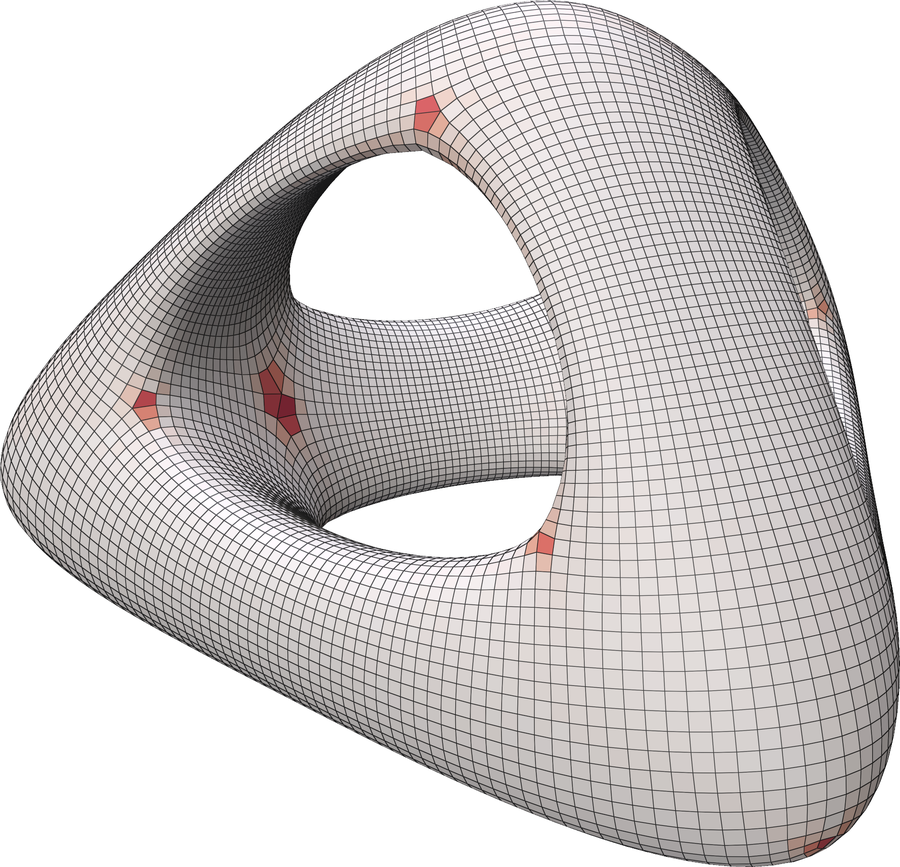} \\
    \end{tabular}
    \caption{Quad mesh results on two smooth models from \cite{Myles:2014:RFG} with area- and angle-distortion minimization.}
    \label{fig:retinal-comparison}
\end{figure}

We demonstrate the efficacy of our $\Phi_\rm{area}$ and $\Phi_\rm{angle}$ distortion energies in \Cref{fig:area-angle-comparison} on two CAD models, M1 and M4 from the MAMBO dataset, and two smooth models from the dataset of \cite{Myles:2014:RFG}, Fertility and Botijo. Two additional smooth models, Retinal and Genus3 are shown in \Cref{fig:retinal-comparison}. The weighting parameters used are those listed above. Clear qualitative differences in the anisotropy of the quads can be seen, and note also that the global singularity positions differ between the two modes of the framework (e.g., on the side of M2 and on Fertility). Another qualitative comment is that one can see rough principal curvature alignment along regions of highest sectional curvature, e.g., the arms of Fertility and Botijo.

Also displayed via colorbars are area and angle distortion metrics. The area disortion measure is the log of the quad area over the target area. As our quads are all nearly planar, we simply measure area by splitting along a diagonal and summing the resulting triangle areas. The angle distortion metric is log of max dimension length over min dimension length. These ``height'' and ``width'' dimension lengths are estimated by averaging the length of opposing sides of the quads. As can be seen, the distortion is mostly centered around singularities, trade-offs that the method deems necessary for reducing distortion more globally on the model.

\begin{figure*}[htbp]
    \centering
    \begin{tabular}{@{}c@{}c@{}c@{}c@{}}
        \includegraphics[width=.25\textwidth]{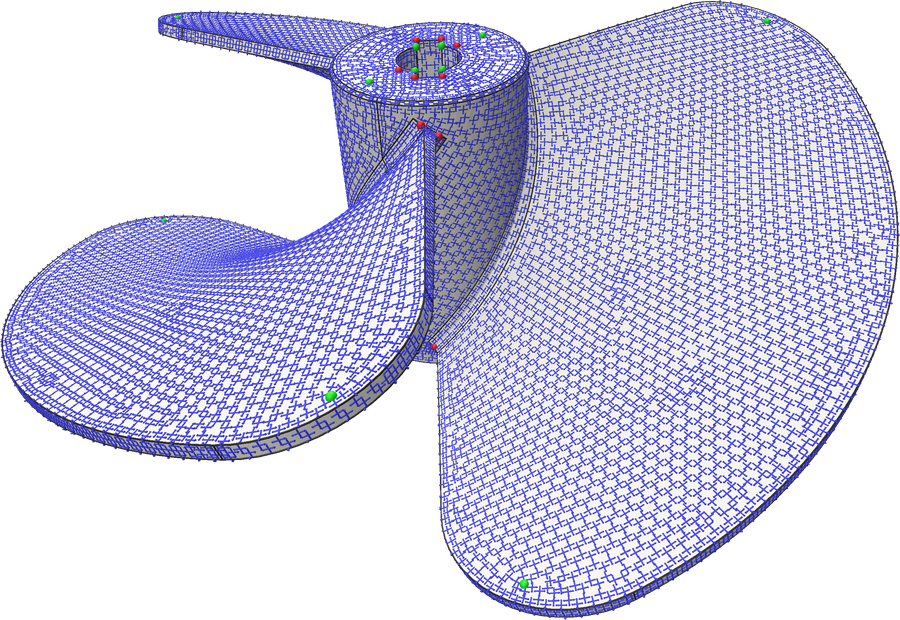}&%
        \includegraphics[width=.25\textwidth]{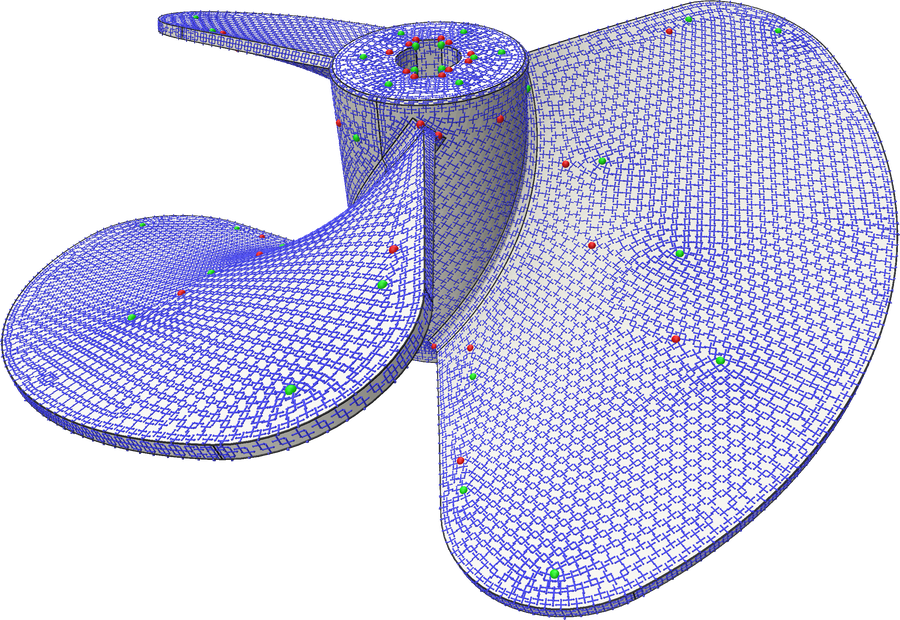}&%
        \includegraphics[width=.25\textwidth]{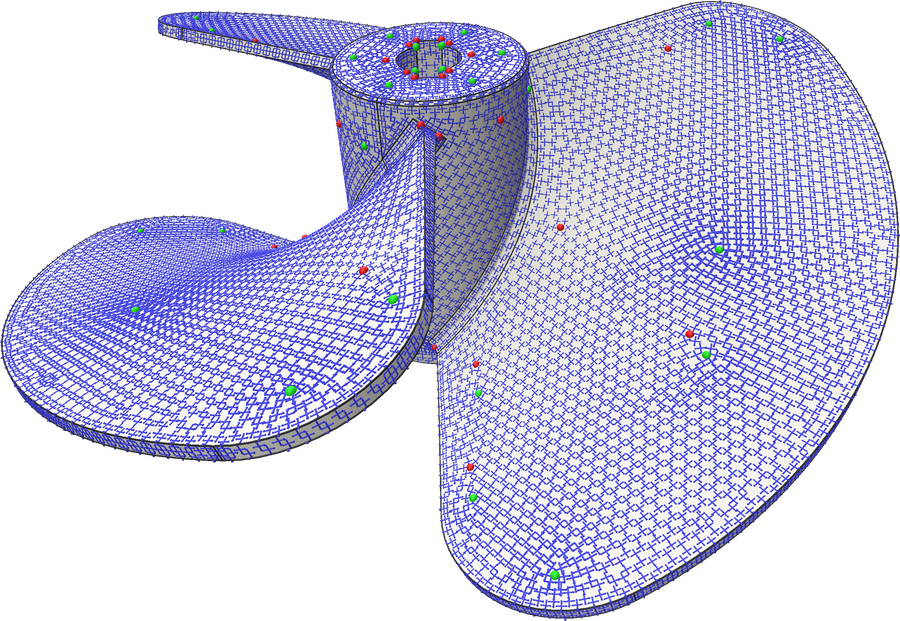}&%
        \includegraphics[width=.25\textwidth]{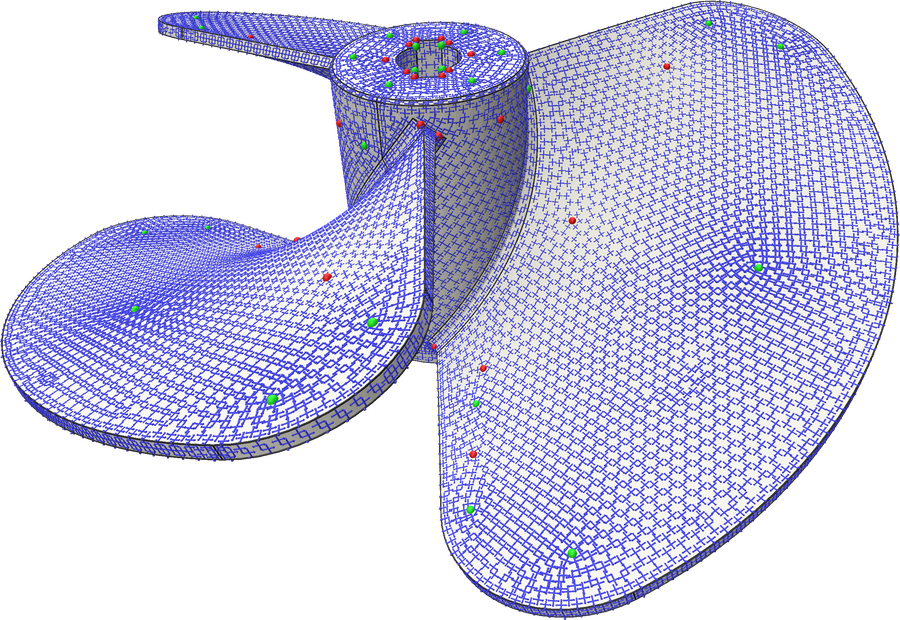}\\
        \small Iter.~0, $n_3=22$, $n_5=22$ &
        \small Iter.~250, $n_3=73$, $n_5=73$ &
        \small Iter.~750, $n_3=64$, $n_5=64$ &
        \small Iter.~1395, $n_3=59$, $n_5=59$ \\
    \end{tabular}
    \caption{
        \revision{
        Evolution of the integrable frame field optimization
        producing the mesh of \cref{fig:teaser} (size is set to 1 along feature curves,
        and we minimize angle distortion)
        on MAMBO model M8.
        Green and red dots represent singularities of valence 3 and 5, respectively.
        Note the balanced singularity counts $n_3$ and $n_5$, which is due to the torus topology
        (Euler characteristic is zero).}
    }
    \label{fig:iters}
\end{figure*}

\revision{
\Cref{fig:iters} shows the evolution of the integrable frame field optimization
(step 2 of the pipeline, \cref{sec:pipeline}),
along with the singularity counts and positions.
The solver introduces and places new singularities to achieve the desired
sizing and distortion objectives.
}

\begin{figure}[htbp]
    \centering
    \def\svgwidth{.8\linewidth}%
\begingroup%
  \makeatletter%
  \providecommand\color[2][]{%
    \errmessage{(Inkscape) Color is used for the text in Inkscape, but the package 'color.sty' is not loaded}%
    \renewcommand\color[2][]{}%
  }%
  \providecommand\transparent[1]{%
    \errmessage{(Inkscape) Transparency is used (non-zero) for the text in Inkscape, but the package 'transparent.sty' is not loaded}%
    \renewcommand\transparent[1]{}%
  }%
  \providecommand\rotatebox[2]{#2}%
  \newcommand*\fsize{\dimexpr\f@size pt\relax}%
  \newcommand*\lineheight[1]{\fontsize{\fsize}{#1\fsize}\selectfont}%
  \ifx\svgwidth\undefined%
    \setlength{\unitlength}{231.06330691bp}%
    \ifx\svgscale\undefined%
      \relax%
    \else%
      \setlength{\unitlength}{\unitlength * \real{\svgscale}}%
    \fi%
  \else%
    \setlength{\unitlength}{\svgwidth}%
  \fi%
  \global\let\svgwidth\undefined%
  \global\let\svgscale\undefined%
  \makeatother%
  \begin{picture}(1,1.18861791)%
    \lineheight{1}%
    \setlength\tabcolsep{0pt}%
    \put(0,0){\includegraphics[width=\unitlength,page=1]{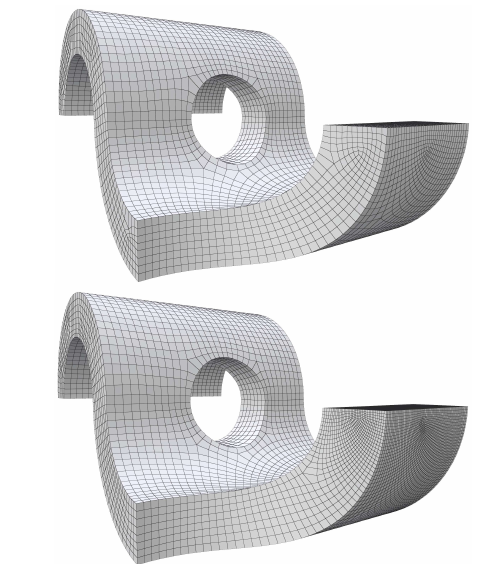}}%
    \put(0.09893811,0.80931544){\color[rgb]{0,0.06666667,0.16862745}\makebox(0,0)[t]{\smash{\begin{tabular}[t]{c}size = 1\end{tabular}}}}%
    \put(0,0){\includegraphics[width=\unitlength,page=2]{size-transition-B1.pdf}}%
    \put(0.09849262,0.21935745){\color[rgb]{0,0.06666667,0.16862745}\makebox(0,0)[t]{\smash{\begin{tabular}[t]{c}size = 1\end{tabular}}}}%
    \put(0,0){\includegraphics[width=\unitlength,page=3]{size-transition-B1.pdf}}%
    \put(0.78933628,1.03175323){\color[rgb]{0,0.06666667,0.16862745}\makebox(0,0)[t]{\smash{\begin{tabular}[t]{c}size = 0.5\end{tabular}}}}%
    \put(0,0){\includegraphics[width=\unitlength,page=4]{size-transition-B1.pdf}}%
    \put(0.77474371,0.44979495){\color[rgb]{0,0.06666667,0.16862745}\makebox(0,0)[t]{\smash{\begin{tabular}[t]{c}size = 0.2\end{tabular}}}}%
    \put(0,0){\includegraphics[width=\unitlength,page=5]{size-transition-B1.pdf}}%
  \end{picture}%
\endgroup%
    \caption{\revision{Example of size transitions with angle distortion minimization
    on MAMBO model B1.
    Sizes are set on the boundaries of the rectangular front and back faces
    designated by the arrows.
    }}
    \label{fig:size-transition}
\end{figure}

\revision{
The size prescription capability allows users to achieve size transitions
in specific regions. This is illustrated in \cref{fig:size-transition}
where we perform a 1-2 and a 1-5 size transition.
The solver inserts pairs of 3-5 singularities to achieve the desired element sizing.
}

\begin{figure}[htbp]
    \centering
    \def\svgwidth{\linewidth}%
\begingroup%
  \makeatletter%
  \providecommand\color[2][]{%
    \errmessage{(Inkscape) Color is used for the text in Inkscape, but the package 'color.sty' is not loaded}%
    \renewcommand\color[2][]{}%
  }%
  \providecommand\transparent[1]{%
    \errmessage{(Inkscape) Transparency is used (non-zero) for the text in Inkscape, but the package 'transparent.sty' is not loaded}%
    \renewcommand\transparent[1]{}%
  }%
  \providecommand\rotatebox[2]{#2}%
  \newcommand*\fsize{\dimexpr\f@size pt\relax}%
  \newcommand*\lineheight[1]{\fontsize{\fsize}{#1\fsize}\selectfont}%
  \ifx\svgwidth\undefined%
    \setlength{\unitlength}{435.61695009bp}%
    \ifx\svgscale\undefined%
      \relax%
    \else%
      \setlength{\unitlength}{\unitlength * \real{\svgscale}}%
    \fi%
  \else%
    \setlength{\unitlength}{\svgwidth}%
  \fi%
  \global\let\svgwidth\undefined%
  \global\let\svgscale\undefined%
  \makeatother%
  \begin{picture}(1,0.63281645)%
    \lineheight{1}%
    \setlength\tabcolsep{0pt}%
    \put(0,0){\includegraphics[width=\unitlength,page=1]{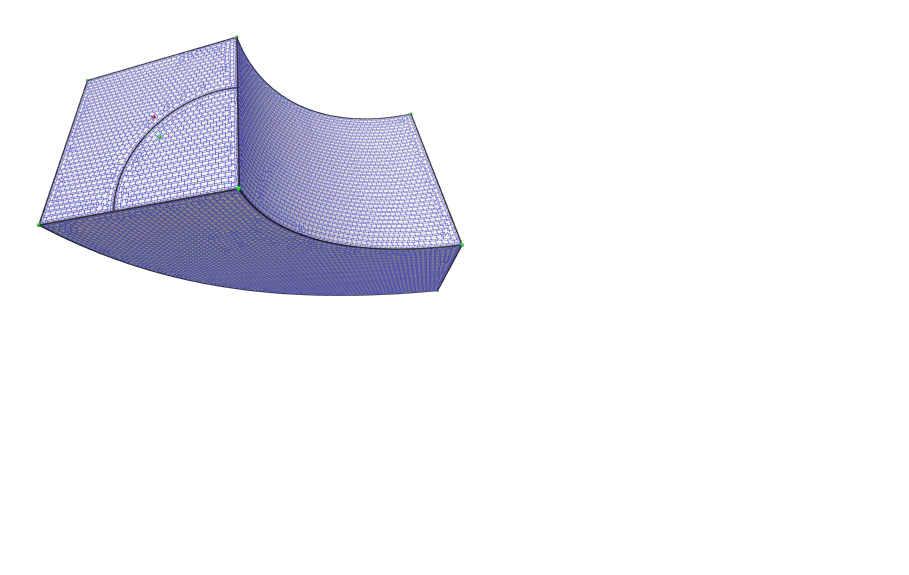}}%
    \put(0.27065836,0.61466873){\color[rgb]{0,0.06666667,0.16862745}\makebox(0,0)[t]{\smash{\begin{tabular}[t]{c}Frame field\end{tabular}}}}%
    \put(0.01675766,0.1496625){\color[rgb]{0,0.06666667,0.16862745}\rotatebox{90}{\makebox(0,0)[t]{\smash{\begin{tabular}[t]{c}Integrable\end{tabular}}}}}%
    \put(0.01728084,0.44926817){\color[rgb]{0,0.06666667,0.16862745}\rotatebox{90}{\makebox(0,0)[t]{\smash{\begin{tabular}[t]{c}Smooth\end{tabular}}}}}%
    \put(0.75545328,0.44027722){\color[rgb]{0,0.06666667,0.16862745}\makebox(0,0)[t]{\smash{\begin{tabular}[t]{c}\textit{non-meshable}\end{tabular}}}}%
    \put(0.75699727,0.61576736){\color[rgb]{0,0.06666667,0.16862745}\makebox(0,0)[t]{\smash{\begin{tabular}[t]{c}Quad mesh\end{tabular}}}}%
    \put(0,0){\includegraphics[width=\unitlength,page=2]{meshable.pdf}}%
  \end{picture}%
\endgroup%
    \caption{\revision{In the presence of feature curves, 
    smooth frame fields can be unmeshable because of limit cycles (top).
    Our solver inserts new singularities that make the frame field meshable (bottom).
    Here area distortion is minimized ($\kappa_\rm{odeco} = 1, \kappa_\rm{area} = 0.1$),
    with no size prescription.}}
    \label{fig:meshable}
\end{figure}

\revision{
In some cases, our integrable solver can take as input a non-meshable
smooth frame field and make it meshable.
\Cref{fig:meshable} shows an example where a smooth frame field
is not meshable; this is due to an invalid frame field topology
where separatrices emanating from singularities form limit cycles.
Our solution (with area distortion minimization) produces new singularities
that make the frame field globally meshable.
}

\subsection{Comparison with Integrable PolyVector Fields}
We compared with the work of \cite{Diamanti2015Integrable}, which we abbreviate IPV. This was the only publicly available alternate approach that jointly optimizes for singularity placement and integrability of a frame field together in the setting of orthogonal anisotropic quad meshes. Their method actually does not require strict orthogonality, but it can be strongly motivated by setting a parameter $s$ close to 1 in their ``geometric order'' term (see Eq. 22 and preceding equations in their work). We set $s=0.9$ and $w_b=10$ (the weighting parameter in front of the geometric order energy) for our comparisons. We used the implementation available as part of Version 2.0.0 of the Directional \cite{Vaxman2016DirectionalSurvey} library with default parameters otherwise. A past version of the library was used as an implementation of IPV is not available as part of the current version.

\revision{
A conceptual difference between IPV and our method
(besides different algebraic representations: complex vs. odeco polynomials)
is that our optimization variables are the odeco polynomial coefficients,
and not the explicit frame vectors.
The benefit of this is that some energies become convex
(e.g. $\Phi_\rm{smooth}$ and $\Phi_\rm{angle}$),
and it obviates the need for explicit objective terms to preserve geometric order of frame vectors.
This conceptual advantage is made possible by the existing theory
of odeco tensors and the description of the variety by quadratic polynomials.
}

\begin{table}[bp]
    \centering
    \caption{Skewness and runtimes comparison with IPV \cite{Diamanti2015Integrable} on ``Medium'' MAMBO CAD models. On average, our method achieves $3.05^\circ$ mean skewness while IPV achieves $5.07^\circ$, and is more robust over this challenging dataset. The improvements come at the price of an average runtime that is $2.17\times$ longer than IPV.}
    \label{tab:ipv-comparison}
    \begin{tabular}{@{}lccrr@{}}
        \toprule
        Model & \multicolumn{2}{c}{Mean Skewness ($^\circ$)} & \multicolumn{2}{c}{Runtime (m:ss)} \\
        \cmidrule(lr){2-3} \cmidrule(l){4-5}
              & IPV & Ours & IPV & Ours \\
        \midrule
        M1 & \ang{4.15} & \ang{3.00} & 4:03 & 11:54 \\
        M2 & \ang{5.78} & \ang{2.39} & 1:58 & 7:31 \\
        M3 & --- & --- & 5:24 & 12:18 \\
        M4 & \ang{3.32} & \ang{3.22} & 3:28 & 7:01 \\
        M5 & \ang{3.13} & \ang{3.29} & 8:08 & 17:14 \\
        M6 & --- & \ang{1.83} & 5:43 & 6:04 \\
        M7 & --- & \ang{4.85} & 11:05 & 9:23 \\
        M8 & \ang{8.98} & \ang{3.58} & 3:24 & 8:58 \\
        M9 & --- & \ang{2.27} & 3:36 & 6:29 \\
        \midrule
        Mean & \ang{5.07} & \ang{3.05} & & \\
        Std & \ang{2.17} & \ang{0.88} & & \\
        \bottomrule
    \end{tabular}
\end{table}

The models considered in our comparison were all of the ``Medium'' models (the hardest of the three categories) in the MAMBO dataset. These CAD models come with feature curves on the boundaries of parametric patches, and we imposed alignment and tangential sizing of 1 on all such curves. No distortion energies were imposed as IPV does not directly accommodate these in their formulation. Over these models, we measured the \emph{skew} of each quad, which is the largest absolute difference from $90^\circ$ of the four corners. Our results, \revision{reported in \cref{tab:ipv-comparison},} were consistently better in terms of orthogonality, with a mean skew of $3.05^\circ$ with standard deviation $0.88^\circ$, while IPV's mean skew was $5.07^\circ$ with standard deviation $2.17^\circ$.

The above averages were calculated over models where the frame field results of both methods could be used to successfully generate a quad mesh (with the procedure outlined in \Cref{sec:mesh_gen}). IPV suffered 4 failures over the 9 models, while our method suffered 1 failure on M3 (shared with IPV). The lack of robustness for both methods is often due to the challenging ``thin corner'' cases that often arise in CAD models, e.g., see \Cref{fig:acute-corner}.

\begin{figure}[tbp]
    \centering
    \begin{tabular}{@{}m{1.5em}@{\hspace{0.5em}}c@{\hspace{0.5em}}c@{}}
        & IPV \cite{Diamanti2015Integrable} & Ours (sizing only) \\
        & \multicolumn{2}{c}{\scriptsize\ang{0}\,\includegraphics[height=1em, valign=m]{figures/colorbar-reds.png}\,\ang{45}} \\[0.5em]
        \multirow{2}{*}{\rotatebox[origin=c]{90}{M4}} &
        \includegraphics[width=0.45\linewidth, valign=m]{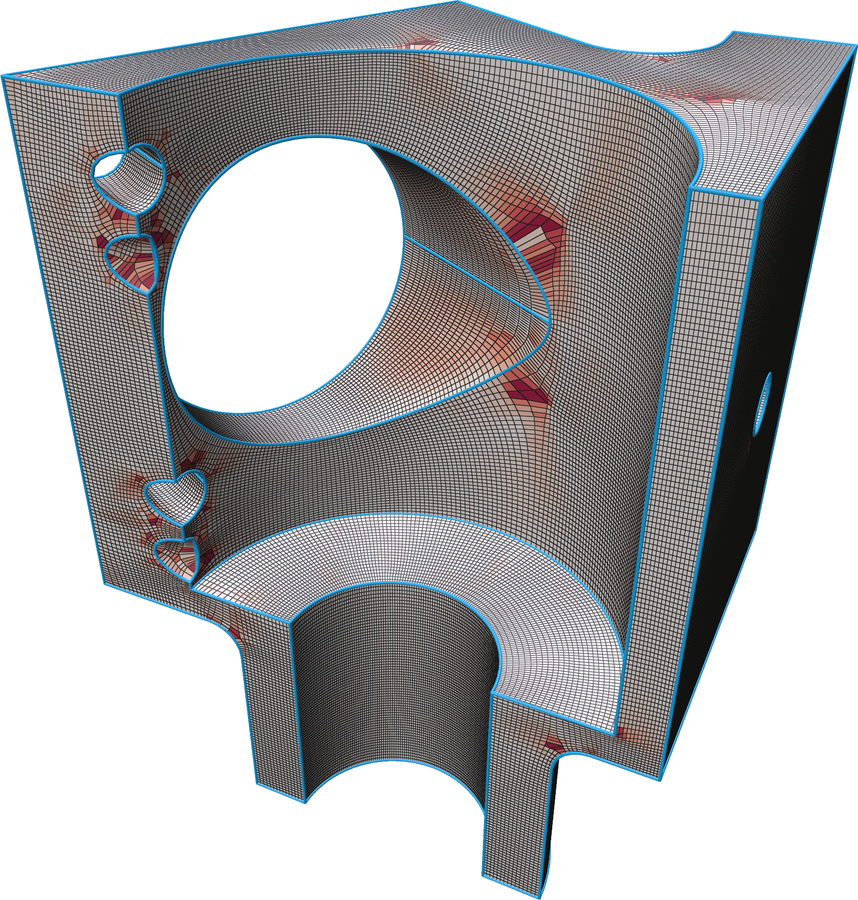} &
        \includegraphics[width=0.45\linewidth, valign=m]{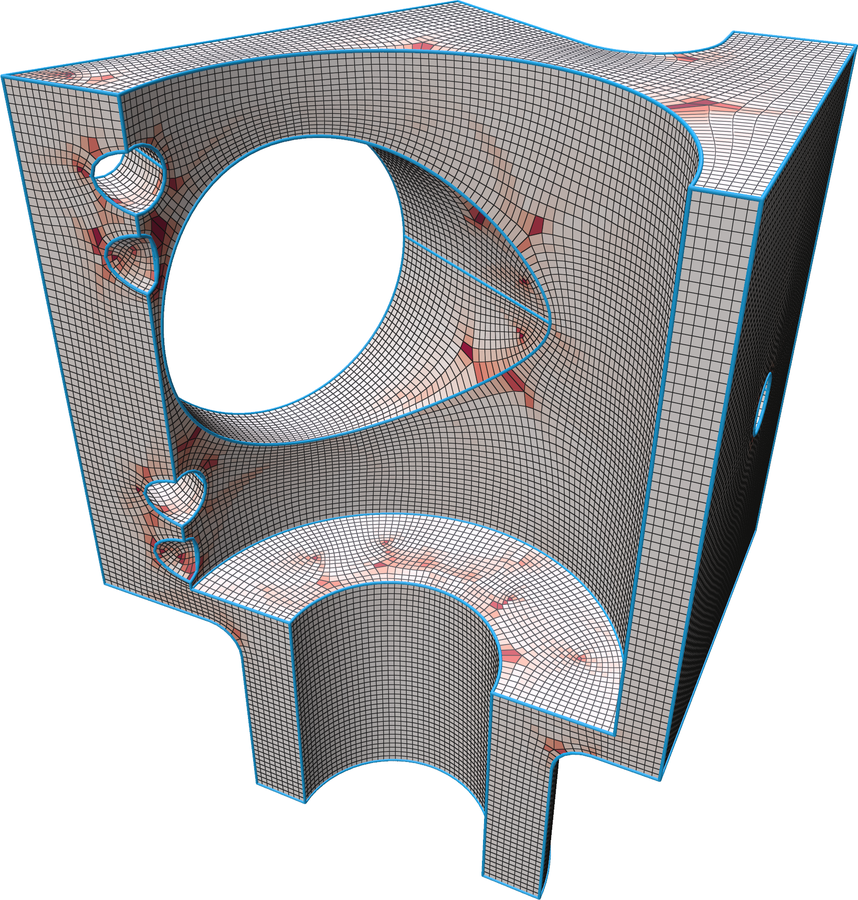} \\
        & \scriptsize(mean skew: \ang{3.32}) & \scriptsize(mean skew: \ang{3.22}) \\[0.5em]
        \multirow{2}{*}{\rotatebox[origin=c]{90}{M8}} &
        \includegraphics[width=0.45\linewidth, valign=m]{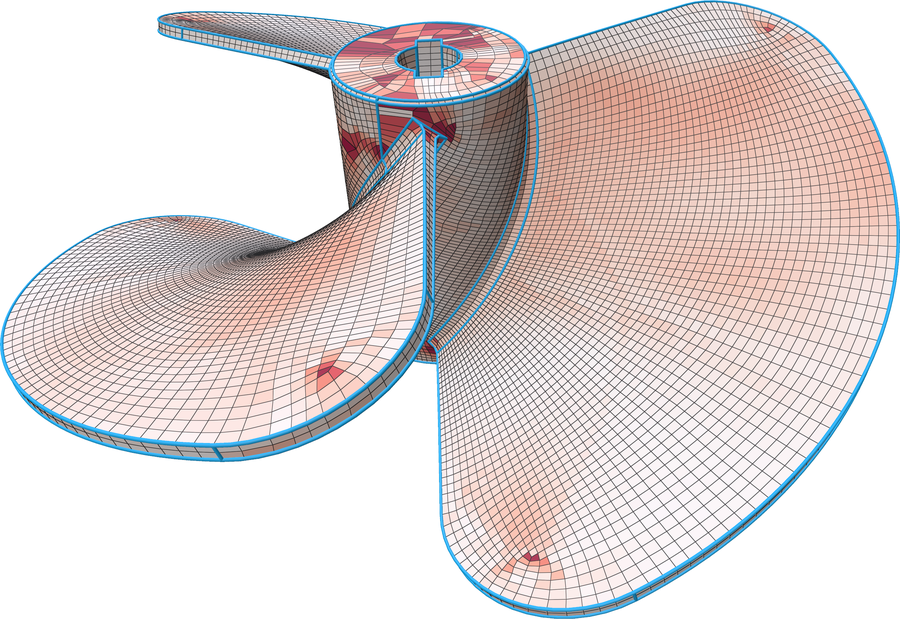} &
        \includegraphics[width=0.45\linewidth, valign=m]{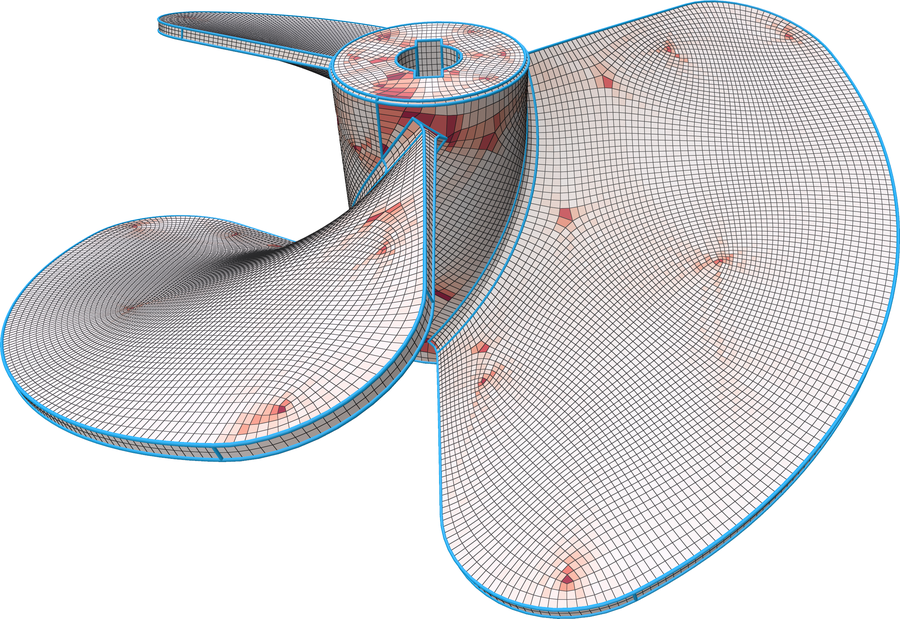} \\
        & \scriptsize(mean skew: \ang{8.98}) & \scriptsize(mean skew: \ang{3.58}) \\
    \end{tabular}
    \caption{A skewness comparison with IPV \cite{Diamanti2015Integrable} on MAMBO models M4 and M8 with uniform tangent sizes on feature curves and no distortion energies. Also note the better satisfaction of sizing constraints via differing singularity configurations,e.g., on the fan blades of M8.}
    \label{fig:polycurl-comparison}
\end{figure}

Results on models M4 and M8 are shown in \Cref{fig:polycurl-comparison}. Our results not only have more orthogonal elements, but also show better satisfaction of boundary sizing constraints. This is due in large part to different singularity configurations, with more numerous pairs of valence 3 and 5 singularities that allow for graceful and effective size transitions. This is quite visible on the fan blades of M8.

\subsection{Comparison with Rectangular Surface Parameterization}

We compare here with the work of \cite{Corman25Rectangular}, which we abbreviate RSP. In particular, we consider the greedy, iterative automatic placement of singularities proposed in Section 5.3.2 of their paper. Without a public implementation, we asked the authors to run their method on four meshing scenarios shown in \Cref{fig:rsp-comparison} utilizing the $\Phi_\rm{iso}$, $\Phi_\rm{area}$, and $\Phi_\rm{angle}$ objectives described in Section 5.1.1 of their work. For a fair comparison of distortion metrics, we compare across a similar number of singularities for each scenario.

The scenarios presented are three challenging planar ones: SquareLine, SquareCurve, and Uturn; and one surface model, S4 from the ``Simple'' (2nd hardest) category from the MAMBO dataset. On all of these models, we impose feature alignment with tangent sizing constraints of 1. The distortion metrics visualized over the meshes are the same as those described in \Cref{sec:distortion_demo} for \Cref{fig:area-angle-comparison}. The reported parenthetical numbers are mean values, derived as follows: for area distortion, we use the log squared of the ratio between quad area and target area; for angle distortion, we use the log squared of the width to height ratios.

Across the planar scenarios, we see that our method produces qualitatively different singularity placements and superior distortion metrics. Also notable are the almost exact symmetry of our singularity placements, relative to those of RSP. In both SquareLine and SquareCurve, one can see differences in positioning relative to the central feature curves and to other singularities in the RSP results, violating the $180^\circ$ rotational symmetry of the scenarios. This is perhaps not unexpected given the greedy, iterative placement strategy of RSP. Lastly, on S4, we compare the use of $\Phi_\rm{iso}$ for their method with weightings of $\kappa_\rm{area} = 0.1$ and $\kappa_\rm{angle} = 0.0001$ (and $\kappa_\rm{odeco} = 10$) for our method. We do not have an explicitly isometry-promoting energy, so we chose this combination of weights. No distortion metrics are visualized on these S4 meshes, but mean values are reported. Perhaps as expected, the results show a clear improvement upon RSP's result in area distortion, and slightly worse performance on angle distortion.

\subsection{Runtimes and Computational Cost} \label{sec:runtimes}

In \Cref{tab:runtimes}, we report the runtimes and mesh statistics for all results that are present in figures in the paper. Our optimization is nonconvex and high-dimensional, with 15 variables per vertex and 27 quadrics being used to define the nonconvex odeco energy. Much of the computational cost derives from these factors, and the time for convergence is naturally hard to predict.

In \Cref{tab:ipv-comparison}, we compare runtimes with IPV on the Medium models of the MAMBO dataset. As can be seen, we are usually 1.5-3x slower, though are sometimes comparable and achieve superior skewness.

\begin{table}[htbp]
    \centering
    \caption{Runtime and mesh statistics for the results shown in this paper.}
    \label{tab:runtimes}
    \begin{tabular}{@{}clccr@{}}
        \toprule
        Fig.  & Model & \#T & \#Q & Runtime (m:ss) \\
        \midrule
        \ref{fig:area-angle-comparison} & M1 (area)       & 37766 & 42279 & 10:57 \\
        & M1 (angle)       & 37766 & 34454 & 3:25 \\
        & M2 (area)       & 22133 & 16942 & 12:06 \\
        & M2 (angle)       & 22133 & 11468 & 8:42 \\
        & Fertility (area) & 27954 & 13352 & 9:13 \\
        & Fertility (angle) & 27954 & 17498 & 4:42 \\
        & Botijo (area) & 82332 & 33550 & 39:23 \\
        & Botijo (angle) & 82332 & 14378 & 26:28 \\
        \midrule
        \ref{fig:retinal-comparison} & Retinal (area) & 7282 & 10906 & 3:58 \\
        & Retinal (angle) & 7282 & 9635 & 2:11 \\
        & Genus3 (area) & 13312 & 19581 & 3:58 \\
        & Genus3 (angle) & 13312 & 17938 & 8:26 \\
        \midrule
        \ref{fig:size-transition} & \revision{B1 (0.5)}      & 12928 & 6869 & 5:38 \\
        & \revision{B1 (0.2)}      & 12928 & 12842 & 6:54 \\
        \midrule
        \ref{fig:meshable} & \revision{B18} & 27749 & 6846 & 13:30 \\
        \midrule
        \ref{fig:polycurl-comparison} & M4      & 38928 & 33324 & 7:01 \\
        & M8      & 26637 & 46492 & 8:58 \\
        \midrule
        \ref{fig:rsp-comparison} & SquareLine (area) & 8678 & 9601 & 3:34 \\
        & SquareLine (angle) & 8678 & 10484 & 4:33 \\
        & SquareCurve (area) & 8775 & 9984 & 1:05 \\
        & SquareCurve (angle) & 8775 & 10444 & 0:47 \\
        & Uturn (area) & 9462 & 5229 & 5:31 \\
        & Uturn (angle) & 9462 & 5765 & 4:55 \\
        & S4 & 30988 & 10180 & 11:26 \\
        \bottomrule
    \end{tabular}
\end{table}

\section{Conclusion}

We present and validate a framework for joint optimization of singularity positions and integrability of a frame field for the purposes of anisotropic surface quad meshing. The framework adapts normal-aligned 3D odeco tensors and nontrivially extends the 2D integrability energy of \cite{Couplet26Size} to that of intrinsic vector curl on embedded, curved surfaces in 3D. We show superior orthogonality and size constraint satisfaction when compared to Integrable PolyVector Fields \cite{Diamanti2015Integrable} on a dataset of challenging CAD models. We also show improved performance with respect to area and angle distortion metrics on a small set of examples when compared to the greedy, iterative automatic placement strategy of \cite{Corman25Rectangular}. Lastly, on a few smooth models from the \cite{Myles:2014:RFG} dataset, we show the efficacy of the method in the absence of features and we see empirical alignment to principal curvature directions in regions of highest sectional curvature. 

\subsection{Limitations and Future Work}

\begin{figure}[h]
    \centering
    \includegraphics[width=0.65\linewidth]{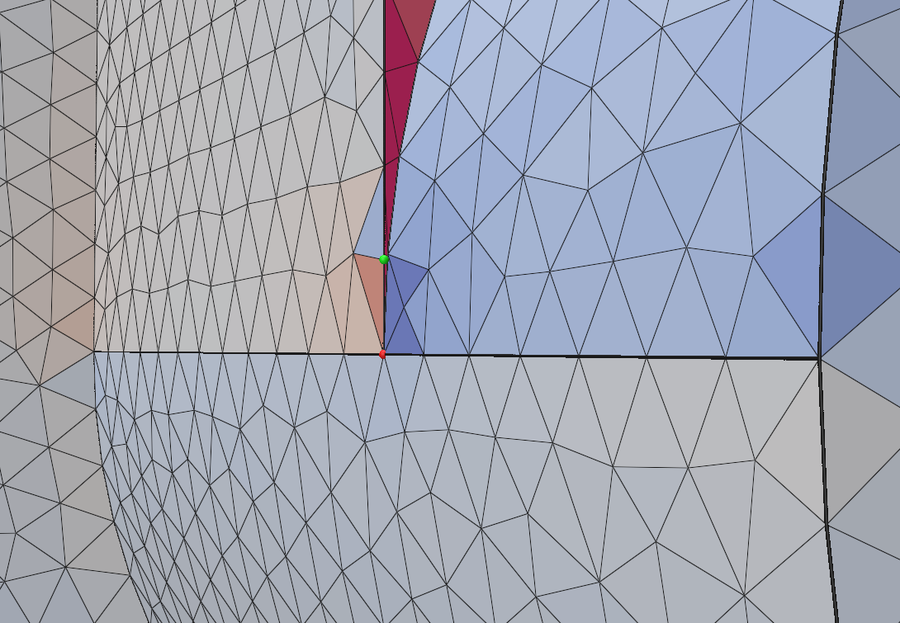}
    \caption{Challenging CAD corners can lead to non-meshable singularity configurations.}
    \label{fig:acute-corner}
\end{figure}

\begin{figure*}[htbp]
    \centering
    \begin{tabular}{@{}m{1.5em}@{\hspace{0.3em}}c@{\hspace{0.3em}}c@{\hspace{0.3em}}c@{\hspace{0.3em}}c@{}}
        & \multicolumn{2}{c}{Minimizing area distortion} & \multicolumn{2}{c}{Minimizing angle distortion} \\[0.5em]
        & RSP \cite{Corman25Rectangular} & Ours & RSP \cite{Corman25Rectangular} & Ours \\[0.5em]
        \multirow{2}{*}[2.4cm]{\rotatebox[origin=c]{90}{SquareLine}} &
        \includegraphics[width=0.23\linewidth]{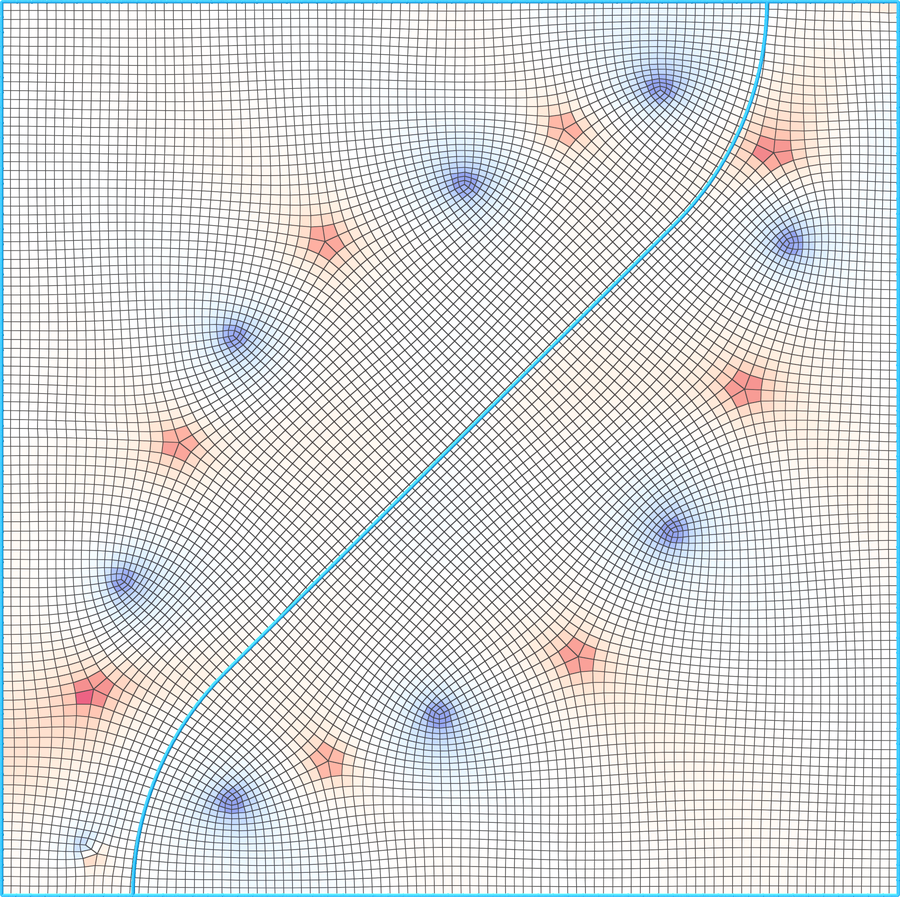} &
        \includegraphics[width=0.23\linewidth]{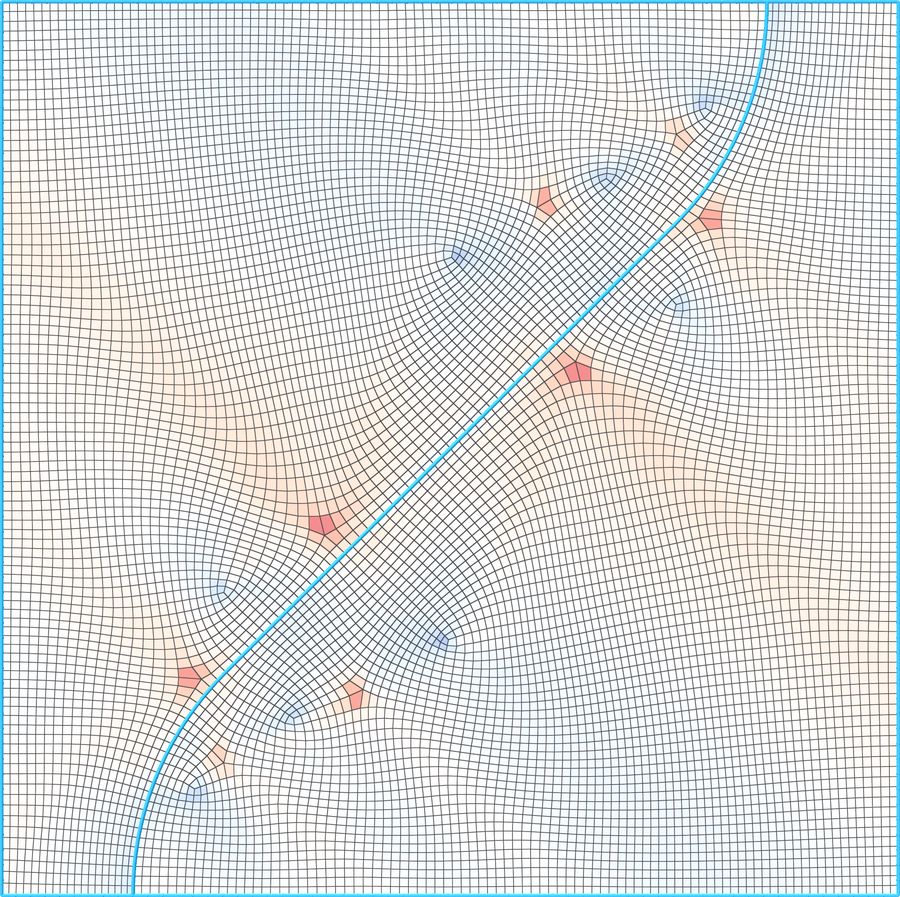} &
        \includegraphics[width=0.23\linewidth]{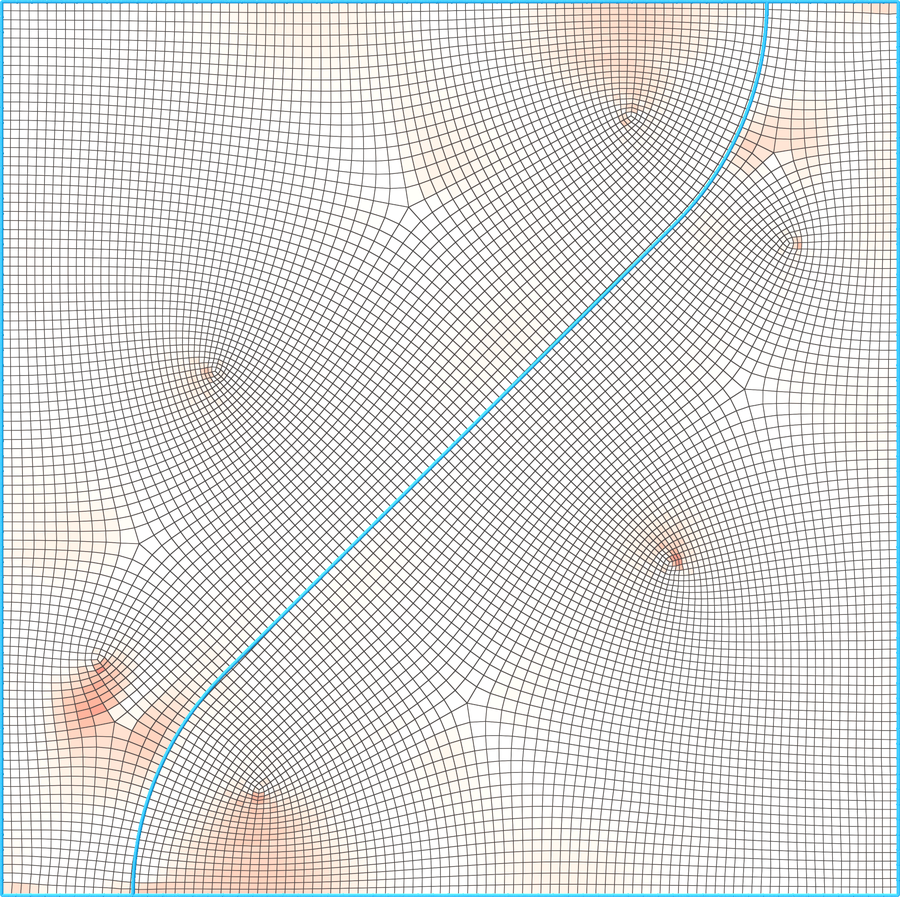} &
        \includegraphics[width=0.23\linewidth]{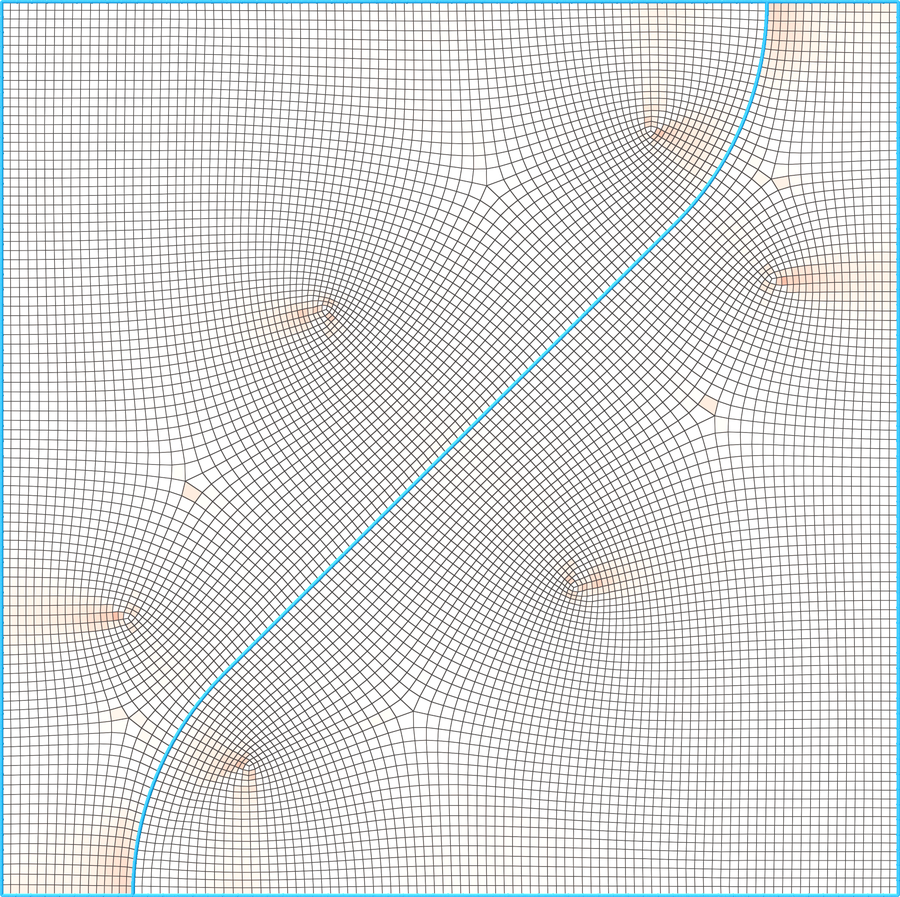} \\
        & \scriptsize(area: 0.047) & \scriptsize(area: 0.022) & \scriptsize(angle: 0.014) & \scriptsize(angle: 0.006) \\[1em]
        \multirow{2}{*}[2.5cm]{\rotatebox[origin=c]{90}{SquareCurve}} &
        \includegraphics[width=0.23\linewidth]{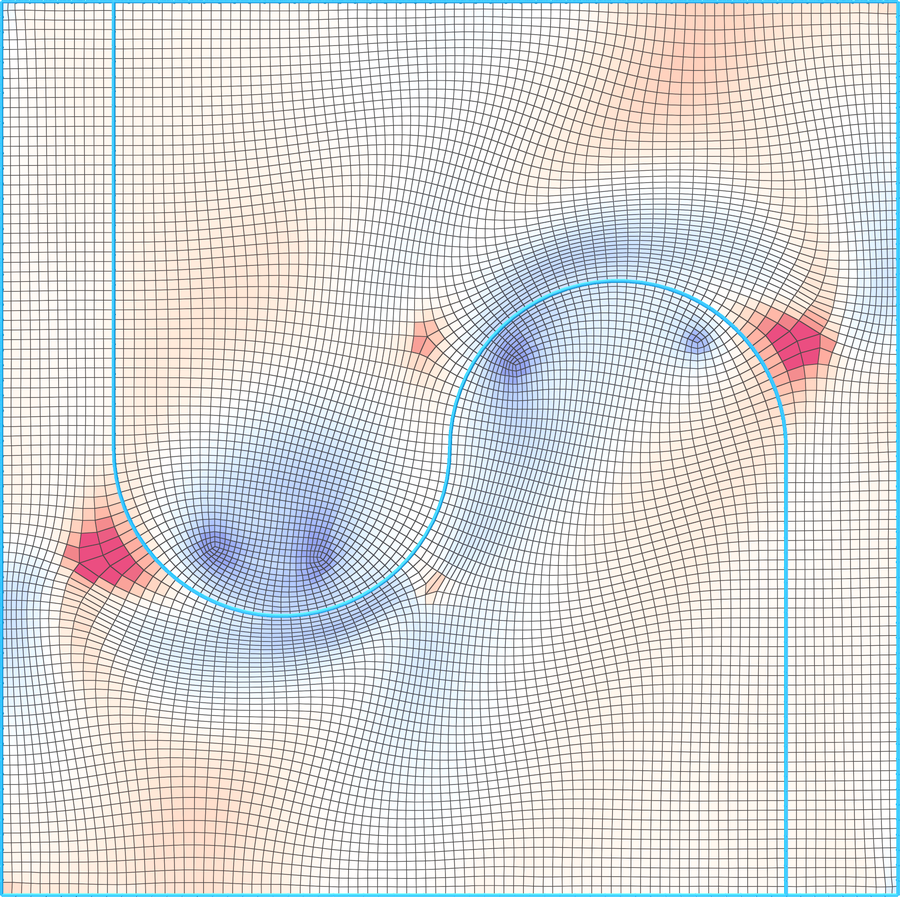} &
        \includegraphics[width=0.23\linewidth]{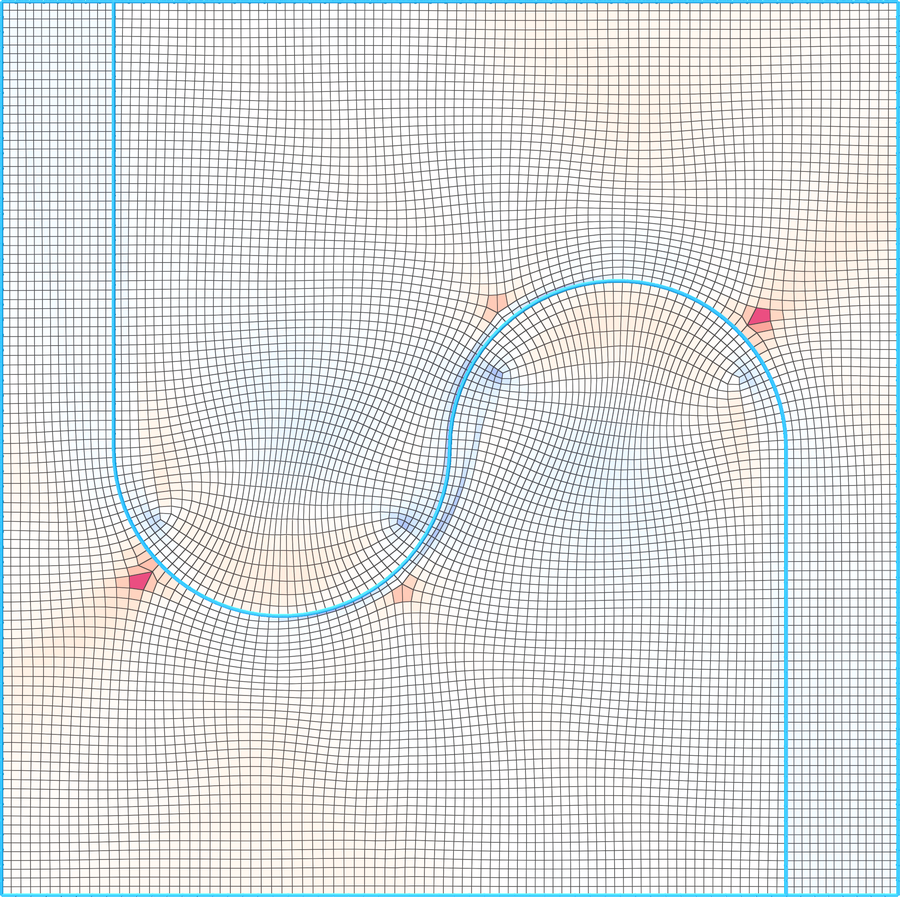} &
        \includegraphics[width=0.23\linewidth]{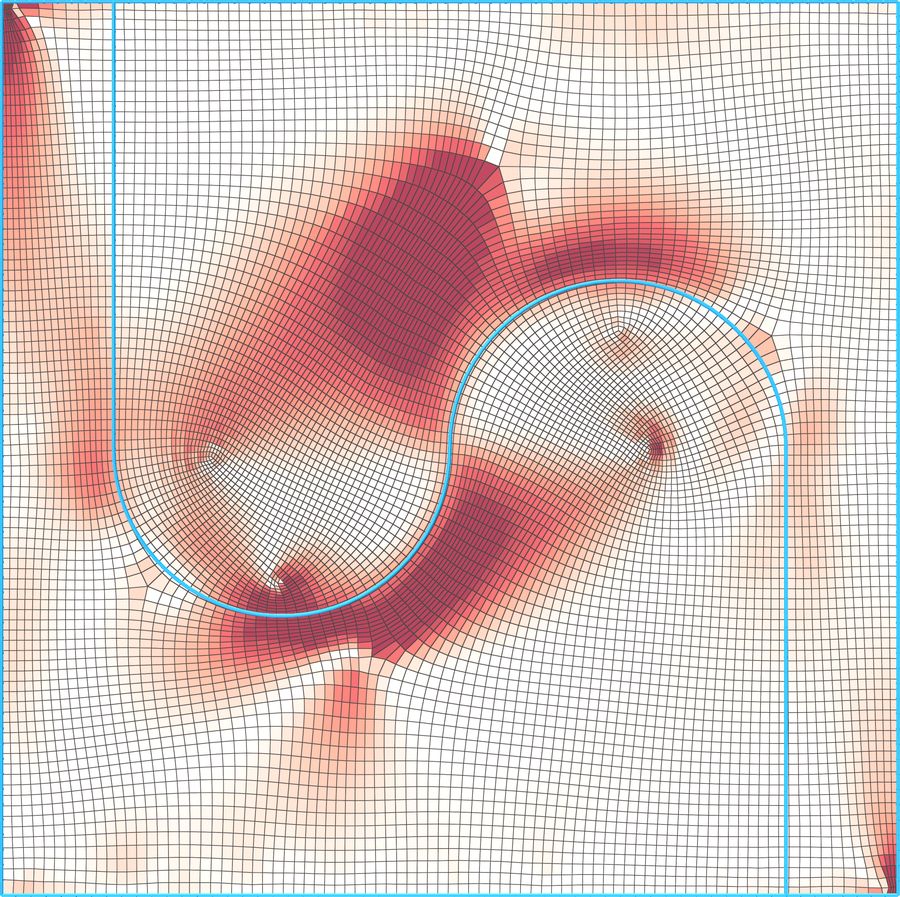} &
        \includegraphics[width=0.23\linewidth]{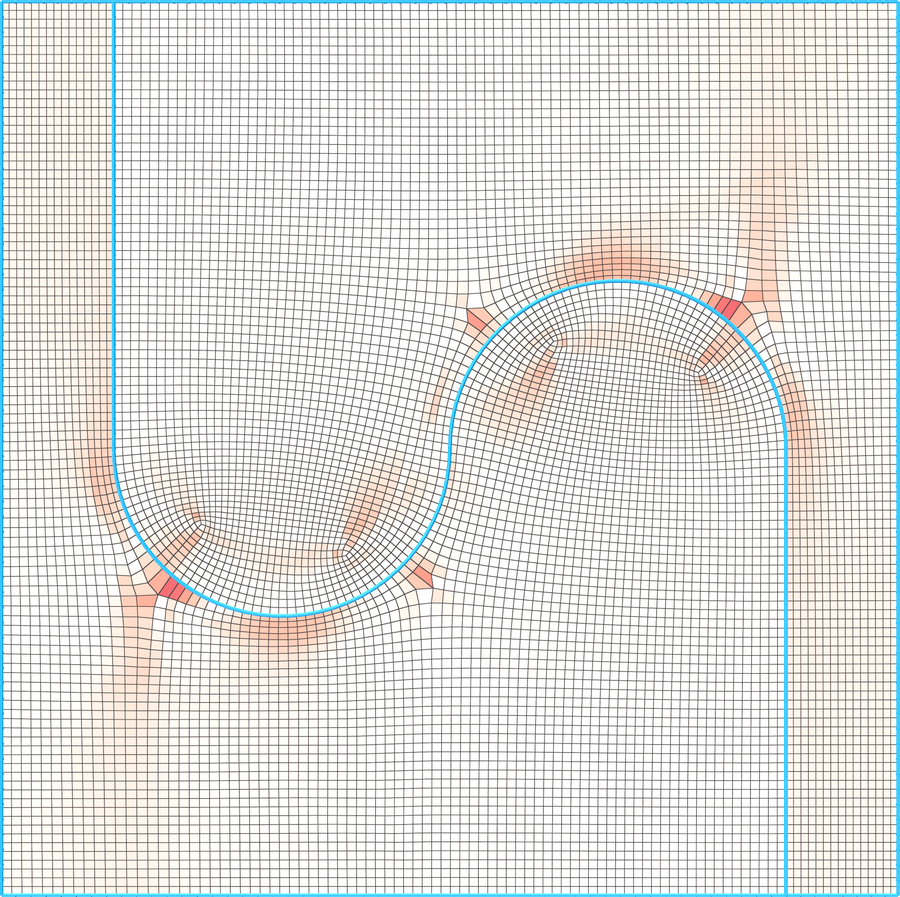} \\
        & \scriptsize(area: 0.104) & \scriptsize(area: 0.017) & \scriptsize(angle: 0.226) & \scriptsize(angle: 0.019) \\[1em]
        \multirow{2}{*}[2.4cm]{\rotatebox[origin=c]{90}{Uturn}} &
        \includegraphics[width=0.18\linewidth]{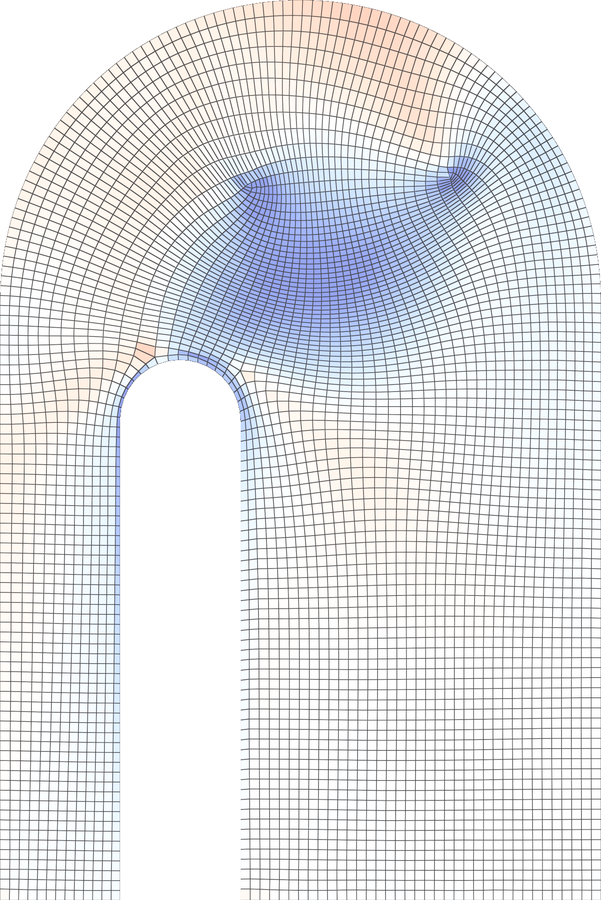} &
        \includegraphics[width=0.18\linewidth]{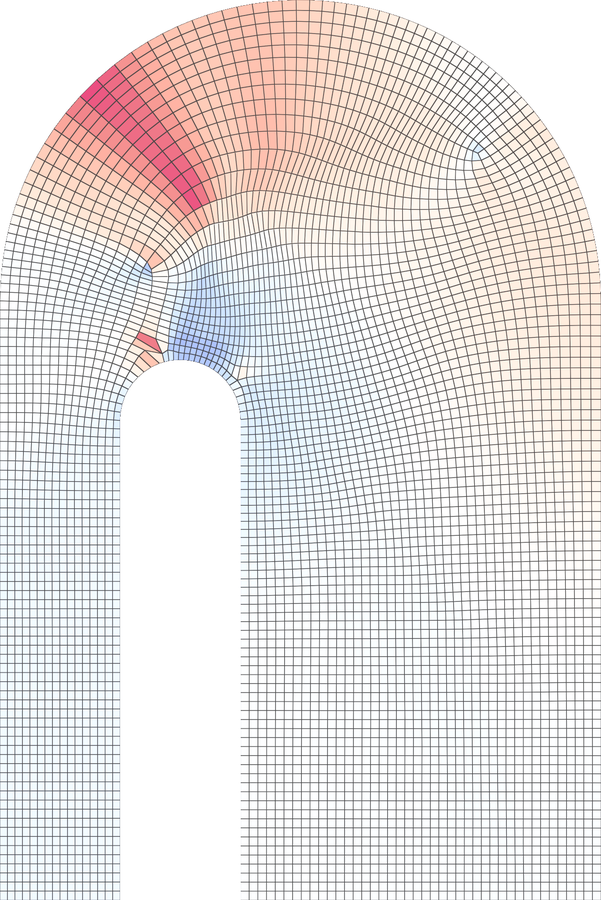} &
        \includegraphics[width=0.18\linewidth]{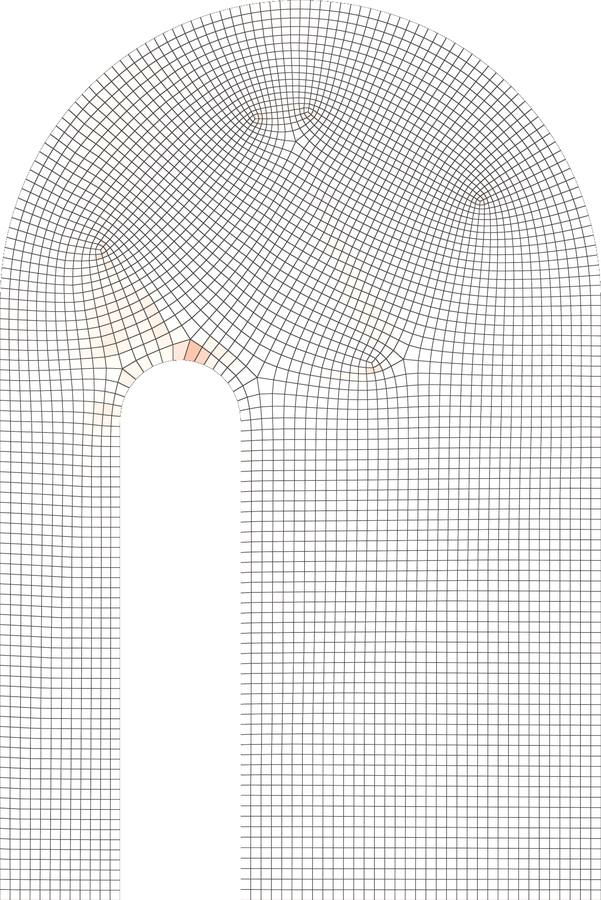} &
        \includegraphics[width=0.18\linewidth]{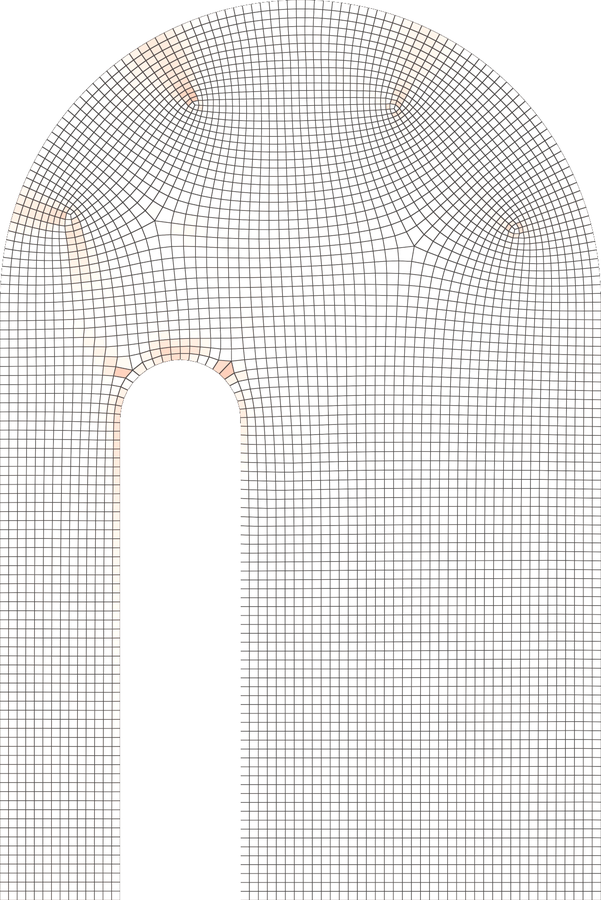} \\
        & \scriptsize(area: 0.154) & \scriptsize(area: 0.061) & \scriptsize(angle: 0.003) & \scriptsize(angle: 0.004) \\[1em]
        \multirow{2}{*}[1.9cm]{\rotatebox[origin=c]{90}{S4}} &
        \multicolumn{2}{c}{\includegraphics[width=0.28\linewidth]{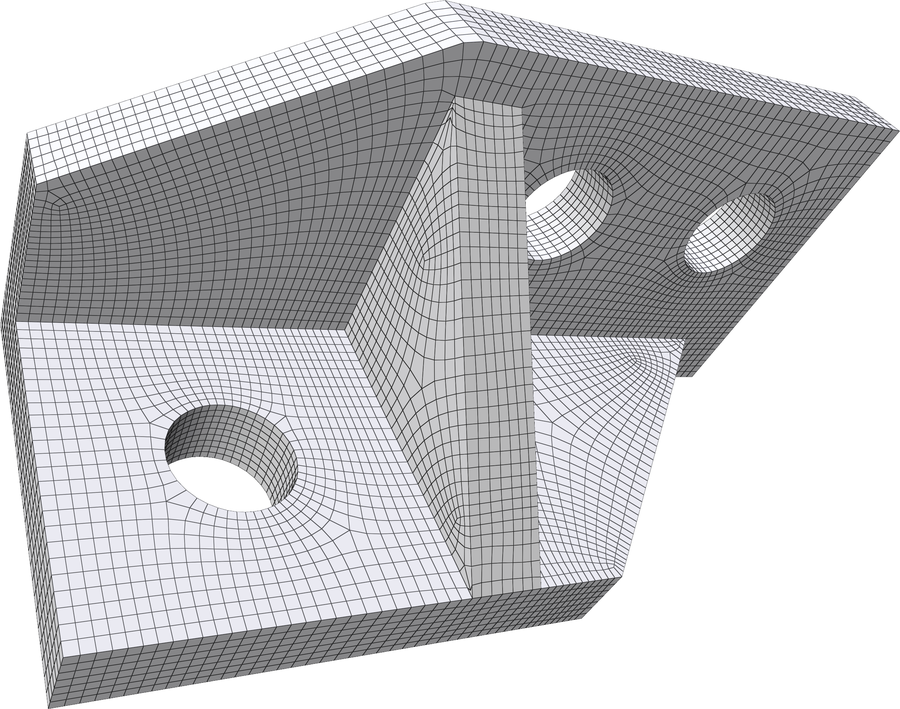}} &
        \multicolumn{2}{c}{\includegraphics[width=0.28\linewidth]{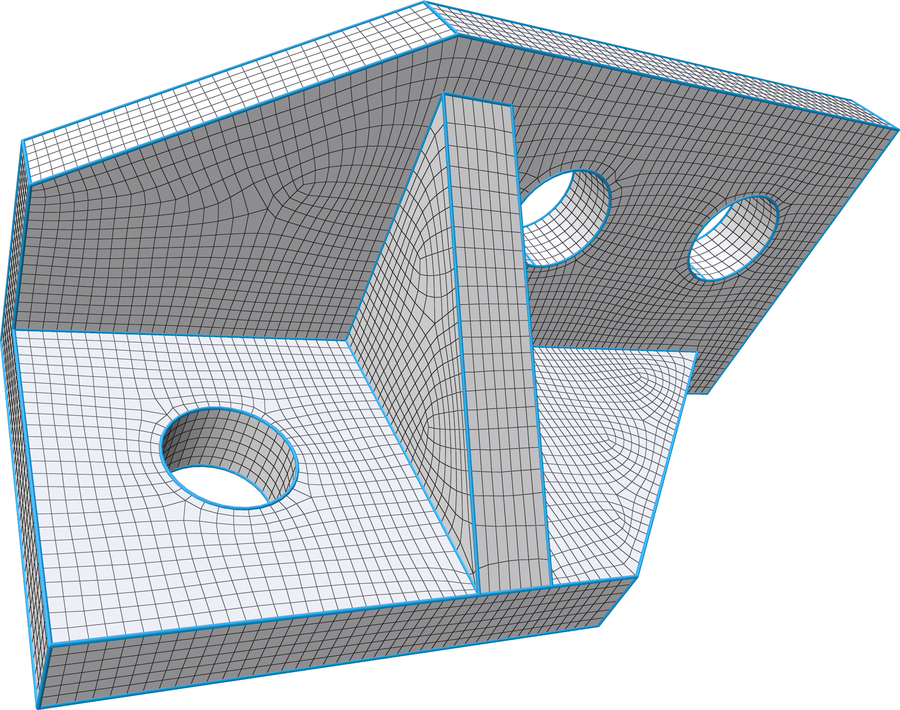}} \\
        & \scriptsize(area: 0.165) & \scriptsize(angle: 0.053) & \scriptsize(area: 0.032) & \scriptsize(angle: 0.077) \\
    \end{tabular}
    \caption{A comparison of minimizing area, angle, and isometric distortion on three challenging planar scenarios and one CAD model. Superior performance at similar singularity counts is seen in the planar models, as well as superior symmetry. On the CAD model, an isometric result for RSP is compared against a mostly area-preserving result from our method.}
    \label{fig:rsp-comparison}
\end{figure*}

Our optimization framework does present several opportunities for improvement and for natural extension of the ideas within. One notable issue is that the method is not fully robust, partially due to the presence of very challenging CAD corners, like the one shown in \cref{fig:acute-corner}. It is likely that a local manual fix at such regions should help to make the method even more robust. Additionally, one could consider feeding our more optimal singularity placements to a method like Rectangular Surface Parameterization \cite{Corman25Rectangular} to even further improve the orthogonality of our end meshes. 

As noted in \Cref{sec:runtimes}, the optimization problem at hand is large and nonconvex, so there is significant computational cost associated with the method. One could attempt alternate optimization strategies, e.g., Gauss-Newton methods, coarser approximation strategies, e.g., one-point quadrature rules, or GPU parallelization of some steps. One notable thought is to attempt an intrinsic odeco tensor approach, dropping the variable count significantly. However, this may lose the extrinsic advantages of natural alignment at sharp creases and regions of high sectional curvature. 

We have also begun preliminary exploration into the exciting prospect of generating integrable volumetric frame fields for hexahedral meshing. The integrability energy is easily adapted to this setting, and it is an enticing prospect to see if the framework can generate reasonable singularity graphs for the analogous parameterization-based hexahedral mesh generation pipeline. 

Lastly, it would be interesting to try to extend the framework to non-orthogonal (non-odeco) frame fields and develop an integrability energy in this setting. The integrability analysis (\Cref{sec:odeco_integrability}) and eigenvalue sensitivity (\Cref{sec:eigensensitivity}) are only valid under the odeco assumption. These could be extended to the non-odeco setting, but the resulting integrability energy becomes more complex: a rational expression of the tensor coefficients.

\section*{Acknowledgments}
We would like to thank Étienne Corman for providing us with additional results from the RSP method.
This work is supported by Wallonie-Bruxelles International and the European Research Council (grant no. 101 071 255).

\FloatBarrier  %
\bibliographystyle{eg-alpha-doi}
\bibliography{manual_bib}

\end{document}